\documentclass[11pt]{article}\pdfoutput=1
\usepackage{cite}
\usepackage{scalerel}
\usepackage{amsmath,amsfonts,amssymb}%,calrsfs}
\usepackage[small,bf,hang]{caption}
\usepackage{slashed}
\usepackage{mathabx,amsmath}
\usepackage{latexsym,epsfig}
\usepackage{dsfont}
\usepackage{arydshln}
\usepackage{extarrows}
\usepackage{hyperref}
\def\hybrid{
        \topmargin -20pt
        \oddsidemargin 0pt
        \headheight 0pt \headsep 0pt
        \textwidth 6.25in % A4 paper
        \textheight 9.5in % A4 paper
        \marginparwidth .875in
        \parskip 5pt plus 1pt \jot = 1.5ex}

\hybrid

 \csname
@addtoreset\endcsname{equation}{section}

\def\moth{\mathsurround=0pt}
\newdimen\zo \zo=0pt

\def\tick{\leaders\hrule height 0.5ex depth 0pt \hskip 0.5pt}
\def\upboxfill{$\moth \setbox\zo\hbox{\tick}%
  \hskip 3pt\hbox to 0pt{$\tick$\hss}\hrulefill \hbox to 7.5pt{$\tick$\hss}$}

\def\dtick{\leaders\hrule height .34pt depth 0.5ex \hskip 0.5pt}
\def\downboxfill{$\moth \setbox\zo\hbox{\dtick}%
  \hskip 2pt\hbox to 0pt{$\dtick$\hss}\hrulefill \hbox to 2pt{$\dtick$\hss}$}

\def\bec{\begin{center}}
\def\ec{\end{center}}

\def\cG{{\cal G}}

\def\cL{{\cal L}}
\def\cD{\mathfrak{D}}
\def\cF{{\cal F}}
\def\cO{{\cal O}}
\def\cA{{\cal A}}

\def\cM{{\cal M}}

\def\cV{{\cal V}}
\def\cY{{\cal Y}}

\def\cA{{\cal A}}

\def\cO{{\cal O}}

\def\del{\partial}

\def\be{\begin{equation}}
\def\ee{\end{equation}}
\def\bea{\begin{eqnarray}}
\def\eea{\end{eqnarray}}
\def\ba{\begin{array}}
\def\ea{\end{array}}

\begin{document}

\begin{titlepage}
\rightline{}
%\rightline\today
\rightline{October 2019}
\rightline{HU-EP-19/27}
\begin{center}
\vskip 1.5cm
 {\Large \bf{Duality Hierarchies and Differential Graded Lie Algebras}}
\vskip 1.7cm

{\large\bf {Roberto Bonezzi and Olaf Hohm}}
\vskip 1.6cm

{\it  Institute for Physics, Humboldt University Berlin,\\
 Zum Gro\ss en Windkanal 6, D-12489 Berlin, Germany}\\
 ohohm@physik.hu-berlin.de\\
roberto.bonezzi@physik.hu-berlin.de 
\vskip .1cm

\vskip .2cm

\end{center}

\bigskip\bigskip
\begin{center} 
\textbf{Abstract}

\end{center} 
\begin{quote}

The gauge theories underlying gauged supergravity and exceptional field theory are 
based on tensor hierarchies: generalizations of Yang-Mills theory
utilizing  algebraic structures that generalize Lie algebras and, as a consequence, 
require higher-form gauge fields. Recently, we proposed that the algebraic structure allowing  
for consistent tensor hierarchies is axiomatized by `infinity-enhanced Leibniz algebras' 
defined on graded vector spaces generalizing Leibniz algebras.  
It was subsequently shown that, upon appending additional vector spaces, 
this structure can be reinterpreted as a differential graded Lie algebra. 
We use this observation to streamline the construction of general tensor hierarchies, 
and we formulate dynamics in terms of a hierarchy of first-order duality relations, 
including scalar fields with a potential.

\end{quote} 
\vfill
\setcounter{footnote}{0}
\end{titlepage}

\tableofcontents

\newpage

\section{Introduction}

In string theory, gauge theories and supergravity it is often instrumental to 
append  to the physical $p-$form gauge fields their on-shell dual $(D-p-2)-$form fields.  
This is necessary, for instance, in order to couple higher branes and to render 
U-duality symmetries manifest and local. Intriguingly, the dynamics is then typically fully encoded 
in the first-order duality relations from which the second-order field equations follow 
as integrability conditions using Bianchi identities.

The theme of higher-form gauge fields is particularly prominent in the U-duality covariant embedding tensor formulation of gauged 
supergravity \cite{deWit:2002vt,deWit:2004nw,deWit:2005hv,deWit:2008ta}. 
Here the gauge algebra structure is no longer a Lie algebra but rather a Leibniz algebra \cite{LODAY,Strobl1,Hohm:2018ybo,Kotov:2018vcz,Lavau:2017tvi}, 
which in turn requires  higher $p-$form gauge fields in order to define gauge covariant 
curvatures. The Bianchi identities then take a hierarchical structure in which the (gauge covariant) 
exterior derivative of a $p-$form field strength is related to the $(p+1)-$form field strength of the next higher form field, 
leading to the notion of tensor hierarchy. 
Consequently, the imposition of first-order duality relations  
requires an entire tower of duality relations, termed \textit{duality hierarchy} in \cite{Bergshoeff:2009ph}, 
from which the second-order equations of gauged supergravity follow as integrability conditions.

Traditionally, the construction of tensor hierarchies of gauged supergravities (and the closely related 
exceptional field theories \cite{Hohm:2013pua,Hohm:2013vpa,Hohm:2013uia,Hohm:2014fxa,Abzalov:2015ega,Musaev:2015ces,Hohm:2015xna,Berman:2015rcc})  
has been done by hand on a case-by-case basis 
up to relatively low form degrees, which gets quickly very tedious. 
Recently, we argued  that the underlying algebraic structure that makes the construction of 
tensor hierarchies to arbitrary degree possible is given by an `infinity-enhanced Leibniz algebra' \cite{Bonezzi:2019ygf}. This algebra is 
defined on a chain complex $X$ with differential ${\frak D}$ and graded symmetric products and extends the 
`enhanced Leibniz algebra' of \cite{Kotov:2018vcz}, which in turn is a (graded) extension of a Leibniz algebra. 
It was pointed out subsequently that upon shifting the grading of $X$ (a step known as suspension) 
and adding additional vector spaces to 
the chain complex $X$, together with an extension of the differential ${\frak D}$, this structure can be 
reinterpreted as a \textit{differential graded Lie algebra} \cite{Lavau:2019oja}. 
In this paper we elaborate on the implications of this observation for the formulation of the tensor hierarchy 
and the tower of duality relations encoding dynamics.

Since there is by now an extensive literature on dualizations in ungauged and gauged supergravity  
let us put the results to be presented here in context. 
In the dimensional reduction of higher-dimensional supergravity (such as $11-$dimensional supergravity) 
one obtains field strengths with Chern-Simons-like modifications, which in turn leads to modified Bianchi identities. 
The resulting dualizations and duality relations have been investigated systematically by Cremmer, Julia, Lu and Pope 
in the seminal papers \cite{Cremmer:1997ct,Cremmer:1998px}. They identified the algebraic structure 
underlying these field strengths, which is \textit{not} a strict Lie algebra but rather an integer graded Lie algebra 
(and thus, in particular, a superalgebra, as it was called in  \cite{Cremmer:1997ct,Cremmer:1998px}). 
These techniques were employed and generalized by Greitz, Howe and Palmkvist in \cite{Greitz:2013pua}, 
who in particular outlined how to include gauged supergravity in terms of differential graded Lie algebras. 
Our investigation was to a large extent inspired by their work.

Before turning to our technical results it is appropriate to discuss  the interplay between  `higher' algebraic structures and more 
conventional Lie type algebras. The infinity-enhanced Leibniz algebra introduced in \cite{Bonezzi:2019ygf}
generalizes Lie algebras in that the Jacobi identity for the bracket defined as the  antisymmetric part of  the  Leibniz product 
does not hold in general. Rather, one obtains an associated $L_{\infty}$ algebra \cite{Zwiebach:1992ie,Lada:1992wc,Lada,Hohm:2017pnh}, 
which in turn implies the need for a hierarchy of higher forms and field strengths. 
On the other hand, the differential graded Lie algebra (dgLa) is a more conventional algebraic structure whose 
brackets satisfy Jacobi identities (albeit graded). There is, however,  nothing paradoxical 
about the fact that a higher algebra can be `derived' in some fashion  from a more conventional algebra. 
In the present case, the Leibniz product $\circ$ is derived from the graded symmetric bracket $[\cdot,\cdot]$ and differential 
${\frak D}$ of the dgLa via $x\circ y=-[{\frak D}x,y]$.\footnote{This is closely related to the `derived bracket' construction of $L_{\infty}$ algebras \cite{Voronov,Getzler}.} 
Thus, to say that the Leibniz product governing the gauge algebra 
is derived from a dgLa does \textit{not} imply that the gauge algebra is secretly a Lie algebra. 
The dgLa construction does imply that the chain complex forms a representation space of a 
genuine Lie algebra $\frak{g}$, which is one of the spaces by which the original complex was extended, 
but $\frak{g}$ plays a somewhat  auxiliary role. More precisely, while 
in gauged supergravity $\frak{g}$  plays the role of the Lie algebra of the global symmetry group of the ungauged limit, 
it is \textit{not} a symmetry of gauged supergravity. 
Moreover, in exceptional field theory it was only quite recently that a Lie algebra $\frak{g}$ playing this role was identified \cite{Hohm:2018ybo,Hohm:2019wql}.

\subsection{Review, Overview and Summary of Results}

Since the results of this paper are quite technical we now briefly review the mathematical background for this investigation and provide a summary 
of our key results. The starting point for a tensor hierarchy is a Leibniz algebra, defined by a generally non-symmetric product $\circ$ satisfying 
 \be\label{LeibnizIDIntro}
  x\circ (y\circ z) - y\circ (x\circ z) \ = \ (x\circ y)\circ z \;. 
 \ee
If this product is antisymmetric the Leibniz algebra reduces to a Lie algebra, but in general it has a symmetric part which we parametrize as
 \be\label{DREL}
  x\circ y +y\circ x \ = \  \cD(x\bullet y)\;. 
 \ee
Here $\bullet$ is a new symmetric operation taking values in a new space that is mapped by the differential $\cD$ back to the Leibniz algebra. 
This relation exhibits the beginning of an entire chain complex $X$ of higher spaces with a differential ${\cD}$ and a \textit{graded symmetric} product $\bullet$, 
satisfying suitable relations that define  what we termed infinity-enhanced Leibniz algebra. 
The forms of the tensor hierarchy then take values in $X$ and can be combined into formal sums of all $p-$forms. 
In particular, the gauge covariant curvatures are combined into a formal sum ${\cal F}$, which satisfy the Bianchi identity 
\begin{equation}\label{oldBianchi00}
D\cF+\tfrac12\,\cF\bullet\cF=\cD\cF\;,    
\end{equation}
where $D$ is the gauge covariant derivative.  Writing out this relation in components, one finds that $D{\cal F}_{p}$ is related to $\frak{D}{\cal F}_{p+1}$, 
thus exhibiting a hierarchical structure. 

The observation of \cite{Lavau:2019oja} was that upon suspension (an overall shift of degree) and upon appending additional vector spaces (including $\frak{g}$ that carries a Lie algebra structure) 
the axioms governing $\bullet$ and $\frak{D}$ take the form of a differential graded Lie algebra (dgLa), where $\bullet$ becomes a (graded) bracket $[\cdot,\cdot]$, 
on which $\cD$ acts as a derivation, and which satisfies a (graded) Jacobi identity.  
We will construct  the corresponding gauge theory by tensoring the space $X$ of the dgLa with the space of forms $\Omega(M)$ on some 
spacetime manifold $M$. Importantly, the resulting space $Z\equiv X\otimes \Omega(M)$ also carries a dgLa structure with respect to a `diagonal' grading. 
The bracket (of degree zero) is defined in terms of the bracket in $X$ and the wedge product of forms, with the differential $\partial$ (of degree -$1$) given by 
 \be
  \partial \equiv d +\cD\;, 
 \ee
where $d$ is the de Rham differential. This operator indeed satisfies $\partial^2=0$ and acts as a derivation on the bracket. 
The gauge fields, gauge parameters and field strengths can again be encoded in formal sums of forms of all degrees 
and are located in the chain complex $Z$ as follows  
\begin{equation}\label{introComplex}
\begin{split}
\cdots 
 {\longrightarrow}\;&Z_1\;\stackrel{\partial}{\longrightarrow}\;Z_0\;\stackrel{\partial}{\longrightarrow}\;Z_{-1}\;
 {\longrightarrow}\;\cdots \;,  \qquad \\
&\Lambda\qquad\quad\,  {\cal A}\qquad\quad\;  {\cal  F}
\end{split}
\end{equation}
where $\Lambda$ denotes the gauge parameters, ${\cal A}$ the gauge fields, and ${\cal F}$ the field strengths. 

The construction of the tensor hierarchy is significantly simplified by the observations in \cite{Greitz:2013pua}, which we employ and improve on 
in this paper. First, the curvatures ${\cal F}$ are closely related to the `pure gauge' object 
 \be\label{OmegaIntro}
  \Omega  \equiv  e^{-{\cal A}}\,\partial \,e^{\cal A}\;, 
 \ee
which identically satisfies the Maurer-Cartan equations 
 \be\label{MCIntro}
  \partial\Omega + \tfrac{1}{2}[\Omega,\Omega]=0\;. 
 \ee
It should be emphasized that the exponential $e^{\cal A}$ of $Z$--valued fields in (\ref{OmegaIntro}) a priori is not well-defined, 
since the dgLa is not defined in terms of associative operators that may be exponentiated, but in the following such structures 
will only appear in combinations like (\ref{OmegaIntro}) where they are interpreted via Baker-Campbell-Hausdorff (BCH)  type formulas. 
The identity (\ref{MCIntro}) can then be verified using only the brackets and graded Jacobi identity of the dgLa. 
Assuming momentarily that there are no scalars, the conventional field strengths ${\cal F}$ can then be extracted from $\Omega$ via 
\begin{equation}\label{OmegaMCdefined0}
\Omega=\cF+\cD A_1\;,    
\end{equation}
where we singled out the  one-form gauge field $A_1$. 
The Maurer-Cartan equation (\ref{MCIntro}) then reads in terms of components, for $p\geq 2$, 
\begin{equation}\label{Bianchicomponent0}
DF_p+\tfrac12\,\sum_{k=2}^{p-1}[F_k,F_{p+1-k}]+\cD F_{p+1}=0 \;,   
\end{equation}
which is precisely equivalent to the Bianchi identity (\ref{oldBianchi00}) of the tensor hierarchy upon suspension. (The resulting sign changes 
will be discussed in detail.) 

Remarkably, the above scheme not only recasts the highly interrelated Bianchi identities of tensor hierarchies 
into the simple form of a Maurer-Cartan equation, but it also allows for the inclusion of scalars 
upon extension of the chain complex by spaces $X_0$, $X_{-1}$, etc.
One may then include zero-forms (scalars) $\phi$ taking values in $\frak{g}=X_0$ by extending the definition of $\Omega$ to 
 \be
  \Omega \ = \ e^{-\phi}e^{-{\cal A}}\,\partial\,\big(e^{{\cal A}} e^{\phi}\big)\;, 
 \ee 
which still satisfies (\ref{MCIntro}).  
Equivalently, we may define the differential 
 \be
  \partial_{\Omega} = \partial + \Omega\;, 
 \ee
viewed as an operator on $X$ that squares to zero, $\partial_{\Omega}^2=0$, as a consequence of the Maurer-Cartan equation (\ref{MCIntro}).
Working this out in terms of components one finds  
\begin{equation}\label{Omegadecomposed0}
\begin{split}
\partial_{\Omega}=D_{Q} + T + P + \sum_{p=2}^{\infty}{\cal V}^{-1}{ F}_{p}{\cal V}\;, 
\end{split}
\end{equation}
with the scalar matrix ${\cal V}\equiv e^{\phi}\in G$ and the `T-tensor' $T\equiv {\cal V}^{-1}\Theta{\cal V}$, where  the `embedding tensor' $\Theta\in X_{-1}$ is defined implicitly  by ${\cal V}^{-1}\cD{\cal V}=[\Theta,\phi]+\cdots$. 
Moreover, we decomposed the $\frak{g}$ valued current ${\cal V}^{-1}D{\cal V}$ into a part, $Q$, taking values 
in a subalgebra $\frak{h}\subset \frak{g}$ and defining the $\frak{h}$--covariant derivatives $D_Q=d+Q$, and its complement $P$. 
Thus, one naturally obtains the structures of non-linear realizations based on $G/H$, and in particular the Maurer-Cartan equation (\ref{MCIntro}) 
encodes then also the Bianchi identities for $P$ and $Q$. Note also that the $A_1$-dependent shift in (\ref{OmegaMCdefined0}) is here absorbed 
into the gauge covariant derivative inside $P$ and $Q$.

As the most intriguing result of this paper, we give  dynamical equations for the fields of the tensor hierarchy, including scalars with a general scalar potential, 
in terms of duality relations. 
To this end we have to assume the existence of $G$--covariant  isomorphisms $I_p:X_p^*\rightarrow X_{n-p-2}$, with $n$ the number of spacetime dimensions and $X^*$ denoting  the dual space, and an $H$--invariant metric $\Delta$, viewed as a collection of maps $\Delta_p: X_p\rightarrow X_p^*$. 
These structures exists in all known examples. In particular, $\Delta$ allows one to extend the `generalized metric' ${\cal M}_1\equiv {\cal V}\Delta_1{\cal V}^{\rm T}$
to a map ${\cal M}$ on the entire chain complex. 
One can then write duality relations for the sum of all curvatures of the form 
\begin{equation}\label{dualitycurved00}
\cF=\star I\cM \cF \;, 
\end{equation}
where $\star$ is the Hodge dual. Using  the Bianchi identities, the integrability conditions for (\ref{dualitycurved00}) imply the second-order equations of motion, 
including the non-linear sigma model equations based on $G/H$ with a source term induced by the scalar potential
 \be
  V = \tfrac{1}{2}(T,\Delta_{-1}T)\;, 
 \ee
where $T\in X_{-1}$ is the T-tensor and round parenthesis indicate the pairing between $X$ and $X^*$. 
In particular, in this formulation the scalar potential is encoded in the $H$--covariant map $\Delta_{-1}$. 
 
The remainder of this paper is organized as follows. 
In sec.~2 we review the observation that the infinity-enhanced Leibniz algebra can be recast as a dgLa. 
In sec.~3 we introduce  the `diagonal complex' encoding the tensor product of this dgLa with the differential forms 
in order to streamline the gauge theory construction of \cite{Bonezzi:2019ygf}, which is 
then extended by including scalars parametrizing a coset space $G/H$. 
Finally, in sec.~4, we explain how to impose duality relations in this formulation from which the second-order field equations 
follow as integrability conditions. We close with a brief outlook in sec.~5, while an appendix contains the explicit field redefinition 
connecting the present formulation  with \cite{Bonezzi:2019ygf}.

\section{Differential graded Lie algebras}

Infinity-enhanced Leibniz algebras have been defined in \cite{Bonezzi:2019ygf} as the algebraic structures that support consistent tensor hierarchies to all orders. Here we will show, reviewing in our language the work of \cite{Lavau:2019oja},  that the product $\bullet$ and the differential $\cD$ define a differential graded Lie algebra upon suspension, \emph{i.e.} degree shifting, of the graded vector space. 

As defined in \cite{Bonezzi:2019ygf}, an infinity-enhanced Leibniz algebra consists of the quadruple $(X,\circ,\cD,\bullet)\,$. $X$ is an $\mathbb{N}-$graded vector space:
\begin{equation}
X=\bigoplus_{n=0}^\infty X_n=X_0+\bar X\;, \end{equation}
where the degree zero subspace $X_0$
is endowed with a (left) Leibniz product $\circ:X_0\otimes X_0\to X_0\,$, obeying
\begin{equation}
x\circ(y\circ z)=(x\circ y)\circ z+y\circ(x\circ z)\;. \end{equation}
$\cD$ is a degree $-1$ differential acting on $\bar X\,$:
\begin{equation}
...\longrightarrow\; X_n\;\stackrel{\cD}{\longrightarrow}\;X_{n-1} \;... \stackrel{\cD}{\longrightarrow}\;X_1\;\stackrel{\cD}{\longrightarrow}\;X_0\;,  \qquad\cD^2 \ = \ 0\;,
\end{equation}
and $\bullet$ is a graded commutative product of degree $+1$ defined on the whole space $X\,$:
\begin{equation}
\bullet:X_i\otimes X_j\to X_{i+j+1}\;,\quad a\bullet b=(-1)^{|a||b|}b\bullet a\;.    
\end{equation}
This quadruple defines an infinity-enhanced Leibniz algebra provided 
\begin{equation}\label{axioms}
\begin{split}
1) \quad &\cD u\circ x=0\;,\quad\forall \;u\in X_1,\; x\in X_0 \;, \\
2)\quad & \cD(x\bullet y)=x\circ y+y\circ x\;,\quad\forall \;x,y\in X_0\;,\\
3)\quad &\cD(x\bullet(y\bullet z))=(x\circ y)\bullet z+(x\circ z)\bullet y-(y\circ z+z\circ y)\bullet x\;,\quad\forall\;x,y,z\in X_0\;,\\
4)\quad &\cD(x_{[1}\bullet(x_{2]}\bullet u))=2\,x_{[2}\bullet\cD(x_{1]}\bullet u)+x_{[2}\bullet(x_{1]}\bullet\cD u)+[x_1,x_2]\bullet u\,,\;\;\forall\,x_1,x_2\in X_0,u\in \bar X\;,\\
5) \quad&\cD(u\bullet v)+\cD u\bullet v+(-1)^{|u|}u\bullet\cD v=0\;,\quad\forall\;u,v\in\bar X\;,\\
6)\quad& (-1)^{|a|}a\bullet(b\bullet c)+(a\bullet b)\bullet c+(-1)^{|b||c|}(a\bullet c)\bullet b=0\;,\quad\forall\;a,b,c\in X\;. 
\end{split}    
\end{equation}
This set of axioms ensures that the generalized Lie derivative
\begin{equation}\label{Lieorigin}
\begin{split}
&\cL_x y:=x\circ y\;,\quad\forall\; x,y\in X_0\;,\\
&\cL_x u:=x\bullet\cD u+\cD(x\bullet u)\;,\quad\forall \;x\in X_0\;, u\in \bar X\;,
\end{split}    
\end{equation}
is covariant w.r.t. $\cD$ and the bullet product, \emph{i.e.}
\begin{equation}
\begin{split}
&[\cL_x,\cD]=0\;,\\
&\cL_x(a\bullet b)=(\cL_x a)\bullet b+a\bullet\cL_x b\;,
\end{split}    
\end{equation}
and that $\cD-$exact degree zero elements generate trivial transformations:
\begin{equation}
\cL_{\cD u}a=0\;,\quad\forall\;u\in X_1\;,\,a\in X\;.    
\end{equation}

\subsection{Suspension}

We will now show that the properties 5) and 6) can be translated into a graded Leibniz property of $\cD$ and graded Jacobi identity in the degree shifted vector space. Let us define the degree shifted vector space $\widetilde X=\bigoplus_{i=1}^\infty\widetilde X_i$  and the suspension $s$ by
\begin{equation}
\begin{split}
&s:X_i\rightarrow \widetilde X_{i+1}  \;,\quad \tilde a:=s a  \;,\\
& |\tilde a|=|a|+1\;.
\end{split}    
\end{equation}
A graded antisymmetric bracket can be defined on $\widetilde X$ by
\begin{equation}\label{dgla bracket}
[\tilde a,\tilde b]:=(-1)^{|a|+1}s(a\bullet b)    \;,
\end{equation}
and by counting degree one can see that $[\,,]$ has intrinsic degree zero. Graded antisymmetry can be proved by
\begin{equation}
\begin{split}
[\tilde b,\tilde a]&= (-1)^{|b|+1}s(b\bullet a)=(-1)^{|a||b|+|b|+1}s(a\bullet b) \\
&=(-1)^{1+|\tilde a||\tilde b|}(-1)^{|a|+1}s(a\bullet b)=(-1)^{1+|\tilde a||\tilde b|}[\tilde a,\tilde b]\;.
\end{split}    
\end{equation}
One can now see that the property 6) above translates to the graded Jacobi identity for the bracket $[\,,]\,$, namely, upon using
\begin{equation}
[[\tilde a,\tilde b],\tilde c]=(-1)^{|a|+|b|}s\big(s^{-1}[\tilde a,\tilde b]\bullet c\big)=(-1)^{|b|+1}s\big((a\bullet b)\bullet c\big)    
\end{equation}
and property 6), one has
\begin{equation}
[[\tilde a,\tilde b],\tilde c]+(-1)^{|\tilde a|(|\tilde b|+|\tilde c|)}[[\tilde b,\tilde c],\tilde a]+(-1)^{|\tilde c|(|\tilde a|+|\tilde b|)}[[\tilde c,\tilde a],\tilde b]=0   \;. 
\end{equation}
This shows that the graded vector space $\widetilde X$ endowed with the bracket $[\,,]$ is a graded Lie algebra.

As for the differential, it can be naturally defined on $\widetilde X$ by (we use the same symbol $\cD\,$, as it should not cause  confusion)
\begin{equation}
\cD\tilde a:= s\,\cD a   \;,
\end{equation}
and retains its degree $-1\,$. From the definition of the bracket \eqref{dgla bracket} one can easily show that property 5) is nothing but compatibility of the differential with the bracket:
\begin{equation}
\cD[\tilde a,\tilde b]=[\cD\tilde a,\tilde b]+(-1)^{|\tilde a|}[\tilde a,\cD\tilde b]\;,\quad |\tilde a|\,,|\tilde b|>1\;,    
\end{equation}
thus establishing that the triple $(\tilde X,[\,,], \cD)$ carries a differential graded Lie algebra (dgLa) structure.

The original Leibniz algebra $(X_0,\circ)$ can be transported to $\widetilde X_1$ by defining\footnote{As with the differential, we denote the new Leibniz product with the same symbol to avoid cluttering formulas.}
\begin{equation}
\tilde x\circ\tilde y :=s(x\circ y)  \;, 
\end{equation}
so that the Leibniz property is unchanged:
\begin{equation}
\tilde x\circ (\tilde y\circ\tilde z)=(\tilde x\circ\tilde y)\circ\tilde z+\tilde y\circ(\tilde x\circ\tilde z) \;. \end{equation}
Notice, however, that the new Leibniz product has intrinsic degree $-1$ and indeed it closes on $\widetilde X_1\,$:
\begin{equation}
\circ: \widetilde X_1\otimes\widetilde X_1\rightarrow\widetilde X_1  \;.  
\end{equation}
At this stage the Leibniz product is the fundamental algebraic structure, to which the dgLa is attached in a compatible way (expressed by axioms 1) to 4) of \eqref{axioms}) in order to construct the full tensor hierarchy. However, it is possible to encode all the axioms \eqref{axioms} in the data of a dgLa, by extending the graded vector space $X$ by a further space $X_{-1}$ of degree $-1$ and allowing in particular to extend $\cD$ as
\begin{equation}
\cD:X_0\rightarrow X_{-1}    \;.
\end{equation}
The role of $X_{-1}$ is naturally interpreted in the degree shifted setting, as the corresponding $\widetilde X_0$ will be a Lie algebra $\frak g\,$, and
\begin{equation}
\cD:\widetilde X_1\rightarrow\frak g    
\end{equation}
plays the role of the usual embedding tensor (see, e.g., \cite{Hohm:2018ybo}). 

\subsection{Adding an extra space}

Having extended the definition of the differential to
\begin{equation}
... \stackrel{\cD}{\longrightarrow}\;X_1\;\stackrel{\cD}{\longrightarrow}\;X_0\;\stackrel{\cD}{\longrightarrow}\;X_{-1}\;,  \qquad\cD^2 \ = \ 0\;,
\end{equation}
one can take the bullet product $\bullet$ and $\cD$ as the primitive structures (hence a dgLa upon suspension), and \emph{define} the Leibniz product as
\begin{equation}\label{Leibniz defined}
x\circ y:=-y\bullet\cD x \equiv-\cD x\bullet y   \;.
\end{equation}
Axioms 5) and 6) of \eqref{axioms} guarantee the Leibniz property:
\begin{equation}
\begin{split}
x\circ (y\circ z)&= -(y\circ z)\bullet\cD x =(z\bullet\cD y)\bullet\cD x  =(z\bullet\cD x)\bullet\cD y-z\bullet(\cD y\bullet\cD x)\\
&=y\circ(x\circ z)+z\bullet\cD(y\bullet\cD x)=(x\circ y)\circ z+y\circ (x\circ z)\;,
\end{split}    
\end{equation}
and all the other axioms 1) to 4) can now be derived by 5) and 6) upon using the definition \eqref{Leibniz defined}. In the degree shifted setting, the Leibniz product \eqref{Leibniz defined} translates into
\begin{equation}\label{Leibniz bracket}
\tilde x\circ \tilde y:=s(x\circ y)=[\tilde y,\cD\tilde x]  \;,  
\end{equation}
and all its properties descend from the dgLa structure. In particular, one can prove the Leibniz identity:
\begin{equation}
\begin{split}
&\tilde x\circ(\tilde y\circ\tilde z)-(\tilde x\circ\tilde y)\circ\tilde z-\tilde y\circ(\tilde x\circ\tilde z) \\
&=[[\tilde z,\cD\tilde y],\cD\tilde x]-[\tilde z,\cD[\tilde y,\cD\tilde x]]-[[\tilde z,\cD\tilde x],\cD\tilde y]\\
&=[[\tilde z,\cD\tilde y],\cD\tilde x]+[[\cD\tilde y,\cD\tilde x],\tilde z]+[[\cD\tilde x,\tilde z],\cD\tilde y]=0\;,
\end{split}    
\end{equation}
by using the graded Jacobi identity and compatibility of the differential. By construction one has $\cD\tilde u\circ\tilde x=0$ from $\cD^2=0\,$, and triviality of the symmetric pairing reads
\begin{equation}
\tilde x\circ\tilde y+\tilde y\circ \tilde x=-\cD[\tilde x,\tilde y]=s\,\cD(x\bullet y)\;.    
\end{equation}
The elements $\cD\tilde x\in\widetilde X_0$ now belong to the degree zero subspace of a dgLa, thus meaning that the space $\widetilde X_0$ itself is a Lie algebra $\frak g\,$. 
Indeed, all the spaces $\widetilde X_n$ carry a representation of the Lie algebra $\widetilde X_0$ induced by the dgLa bracket:
For any Lie algebra element $\tilde\xi\in\widetilde X_0$
one can define $\rho_{\tilde\xi}:\widetilde X_n\rightarrow\widetilde X_n$ by
\begin{equation}
\rho_{\tilde\xi}(\tilde v):=[\tilde\xi,\tilde v]\;,\quad \tilde v\in \widetilde X_n  \;,  
\end{equation}
that is a representation thanks to the graded Jacobi identity, \emph{i.e.}
\begin{equation}
\rho_{\tilde\xi_1}(\rho_{\tilde\xi_2}(\tilde v))-\rho_{\tilde\xi_2}(\rho_{\tilde\xi_1}(\tilde v))=\rho_{[\tilde\xi_1,\tilde\xi_2]}(\tilde v)  \;.  
\end{equation}
The Leibniz product $\tilde x\circ\tilde y$ of two elements in $\widetilde X_1$ has then the form
\begin{equation}
\tilde x\circ \tilde y=-\rho_{\cD\tilde x}(\tilde y)\;,  \end{equation}
that is precisely the usual expression in terms of the embedding tensor $\vartheta$ upon identifying it with $-\cD:\widetilde X_1\rightarrow \frak g\,$, and $\widetilde X_1$ with a $\frak g-$representation $R\,$.

Before suspension one has the same structure upon identifying $g\equiv X_{-1}\,$, with the representation given by
\begin{equation}
\rho_\xi(v):=\xi\bullet v    \;,\quad \xi\in X_{-1}\;,\quad v\in X_n\;.
\end{equation}
According to \eqref{Leibniz defined}, the Leibniz product is indeed
\begin{equation}
x\circ y=-\rho_{\cD x}(y)  \;,  
\end{equation}
and the relation between the two pictures is given by
\begin{equation}
\rho_{\tilde\xi}(\tilde v)=s\big(\rho_\xi(v)\big)\;.    
\end{equation}

The Lie derivative \eqref{Lieorigin} can now be defined universally on $X$ by
\begin{equation}\label{LieLeibnew}
\cL_x a:=-\cD x\bullet a=(-1)^{|a|+1}a\bullet\cD x\;,\quad x\in X_0\;,\;a\in X    \;,
\end{equation}
and it coincides with the previous definition thanks to \eqref{Leibniz defined}. Upon suspension, it can be defined by
\begin{equation}\label{LiedgLa}
\cL_{\tilde x}\tilde a := s(\cL_x a)=[\tilde a,\cD\tilde x] \;,  
\end{equation}
and it retains intrinsic degree zero, as well as its covariance properties:
\begin{equation}
\begin{split}
\cD(\cL_{\tilde x}\tilde a)&=\cD[\tilde a,\cD\tilde x]=[\cD\tilde a,\cD\tilde x]=\cL_{\tilde x}\,(\cD\tilde a)\;,\\[2mm]
\cL_{\tilde x}[\tilde a,\tilde b]&=[[\tilde a,\tilde b],\cD\tilde x]=(-1)^{1+|\tilde a||\tilde b|}[[\tilde b,\cD\tilde x],\tilde a]-[[\cD\tilde x,\tilde a],\tilde b]\\
&=[\cL_{\tilde x}\tilde a,\tilde b]+(-1)^{1+|\tilde a||\tilde b|}[\cL_{\tilde x}\tilde b,\tilde a]=[\cL_{\tilde x}\tilde a,\tilde b]+[\tilde a,\cL_{\tilde x}\tilde b]\;.
\end{split}
\end{equation}
Since from now on we will only work in the suspended dgLa picture, we will drop tildes from all expressions, so that $X=\oplus_{k=0}^\infty X_k$ is the graded Lie algebra with $X_0\equiv\mathfrak{g}\,$, the Leibniz product reads $x\circ y=-[\cD x,y]$ and so on.

\section{Diagonal complex and dgLa gauge theory}

In order to construct a gauge theory based on the above algebraic structure, we introduce the space $\Omega(M)$ of differential forms on a spacetime manifold $M\,$, and tensor it with the dgLa $X\,$, defining $Z:=\Omega(M)\otimes X$ \emph{i.e.} the space of ${X}-$valued differential forms. The space $Z$ naturally inherits the structure of a bi-complex, with bi-grading given by the dgLa degree and form degree. The two differentials are the dgLa one $\cD$ and the de Rham differential $d$ in spacetime. Field strengths, gauge fields, gauge parameters, trivial parameters etc. organize themselves in the bi-complex as shown in the diagram below:

\be\label{DIAGRAM}
\begin{array}{ccccccccccc}\cdots&\xlongrightarrow{\cD} &Z_{[0]}^3 {}  &\xlongrightarrow{\cD}
&Z_{[0]}^2 (\chi_0)&
\xlongrightarrow{\cD}&Z^1_{[0]} (\lambda_0)
\\[1.5ex]
\Big{\downarrow}{{\rm d}}&&\Big{\downarrow}{{\rm d}}&&\Big{\downarrow}{{\rm d}}&&\Big{\downarrow}{{\rm d}}
\\[1.5ex]
\cdots&\xlongrightarrow{\cD}&Z^3_{[1]}(\chi_1) &\xlongrightarrow{\cD}&Z^2_{[1]}(\lambda_1)&
\xlongrightarrow{\cD}&Z^1_{[1]}(A_1)\\[1.5ex]
\Big{\downarrow}{{\rm d}}&&\Big{\downarrow}{{\rm d}}&&\Big{\downarrow}{{\rm d}}&&\Big{\downarrow}{{\rm d}}
\\[1.5ex]
\cdots&\xlongrightarrow{\cD}&Z^3_{[2]}(\lambda_{2}) &\xlongrightarrow{\cD}&Z^2_{[2]}( A_{2})&
\xlongrightarrow{\cD}&Z^1_{[2]}({F}_{2})\\[1.5ex]
\Big{\downarrow}{{\rm d}}&&\Big{\downarrow}{{\rm d}}&&\Big{\downarrow}{{\rm d}}&&\Big{\downarrow}{{\rm d}}
\\[1.5ex]
\cdots&\xlongrightarrow{\cD}&Z^3_{[3]}( A_3) &\xlongrightarrow{\cD}&Z^2_{[3]} ({F}_3)&
\xlongrightarrow{\cD}&Z^1_{[3]} ({\rm d}{ F}_2)
\end{array}
\ee
where the subscript denotes form degree and the superscript dgLa degree. The $A$'s in the diagram are gauge fields, $F$'s are curvatures, $\lambda$'s gauge parameters and so on.
The physical interpretation of the various fields relates their internal (dgLa) degree to the form degree: For instance, any $p-$form gauge field $A_p$ has internal degree $p\,$, curvatures $F_p$ have degree $p-1$ and so on. This suggests that a better way to organize the bi-complex is by choosing a diagonal total degree, given by the difference of the internal and form degrees, already used in the $L_\infty$ construction in \cite{Bonezzi:2019ygf}. The diagonal complex that emerges also carries a dgLa structure, as we will now show.

In order to deal efficiently with sign factors, we introduce odd oscillator variables $\theta^\mu$ that carry intrinsic degree $-1\,$. These can be viewed as the usual $dx^\mu$ with additional commutation properties that help determining the phase factors in the dgLa on the total space $Z$. More precisely, they obey
\begin{equation}\label{thetology}
\theta^\mu\theta^\nu+\theta^\nu\theta^\mu=0\;,\quad\theta^\mu\cD+\cD\theta^\mu=0\;,\quad \theta^\mu\,\omega_{\nu_1...\nu_p}=(-1)^{\alpha_p}\omega_{\nu_1...\nu_p}\,\theta^\mu   \;, 
\end{equation}
where $\alpha_p$ is the internal degree of $\omega_{\nu_1...\nu_p}\,$.
A differential form $\omega_p$ will be written as
\begin{equation}
\omega_p=\tfrac{1}{p!}\,\theta^{\mu_1}...\theta^{\mu_p}\,\omega_{\mu_1...\mu_p}    \;, 
\end{equation}
and we can naturally define the bracket and the action of $\cD$ on the total space by\footnote{In principle, one does not need to introduce the $\theta$ oscillators since one may define differential forms as usual but postulate signs in the definitions of $\cD$ and $[\cdot,\cdot]$: $\cD\omega_p:=\tfrac{(-1)^p}{p!}\,\cD\omega_{\mu_1...\mu_p}\,dx^{\mu_1}\wedge...\wedge dx^{\mu_p}$ and $[\omega_p,\eta_q]:=\tfrac{1}{p!q!}(-1)^{q\alpha_p}[\omega_{\mu_1...\mu_p},\eta_{\nu_1...\nu_q}]dx^{\mu_1}\wedge...\wedge dx^{\mu_p}\wedge dx^{\nu_1}...\wedge dx^{\nu_q}\,$. The use of graded $\theta$ oscillators, however, makes it easier to keep track of the correct sign factors.
}
\begin{equation}\label{totalbracket}
\begin{split}
\cD\omega_p&:=\cD\Big(\tfrac{1}{p!}\,\theta^{\mu_1}...\theta^{\mu_p}\,\omega_{\mu_1...\mu_p}\Big)=\tfrac{(-1)^p}{p!}\,\theta^{\mu_1}...\theta^{\mu_p}\,\cD\omega_{\mu_1...\mu_p} \;,\\
[\omega_p,\eta_q]&:=\tfrac{1}{p!q!}[\theta^{\mu_1}...\theta^{\mu_p}\,\omega_{\mu_1...\mu_p},\theta^{\nu_1}...\theta^{\nu_q}\,\eta_{\nu_1...\nu_q}]\\
&=\tfrac{1}{p!q!}(-1)^{q\alpha_p}\theta^{\mu_1}...\theta^{\mu_p}\theta^{\nu_1}...\theta^{\nu_q}[\omega_{\mu_1...\mu_p},\eta_{\nu_1...\nu_q}]\;,
\end{split}    
\end{equation}
where we denote by $\alpha$ the internal degree of forms, and we used the properties \eqref{thetology} to extract the $\theta$'s to the left.
The de Rham differential takes the form $d=\theta^\mu\del_\mu\,$ and thus carries degree $-1$ and anticommutes with $\cD$ due to \eqref{thetology}: $\{d,\cD\}=0\,$. Its action on differential forms is the usual one:
\begin{equation}
d\omega_p:=\tfrac{1}{p!}\,\theta^{\mu_1}...\theta^{\mu_{p+1}}\,\del_{\mu_1}\omega_{\mu_2...\mu_{p+1}}  \;.  
\end{equation}
Since now both $\cD$ and $d$ have degree $-1$ and anticommute, it is possible to define a total differential:
\begin{equation}
\del:=d+\cD\;,\quad \del^2=0 \;.   
\end{equation}
The advantage of this construction is that the bracket and the differentials have a dgLa structure w.r.t. the diagonal degree defined as $N_p:=\alpha_p-p$ for a $p-$form of internal degree $\alpha_p\,$.
One can see this from the definitions:
\begin{equation}
\begin{split}
[\omega_p,\eta_q]&=\tfrac{1}{p!q!}[\theta^{\mu[p]}\omega_{\mu[p]},\theta^{\nu[q]}\eta_{\nu[q]}] =   \tfrac{1}{p!q!}(-1)^{q\alpha_p}\theta^{\mu[p]}\theta^{\nu[q]}[\omega_{\mu[p]},\eta_{\nu[q]}]\\
&=\tfrac{1}{p!q!}(-1)^{1+\alpha_p\alpha_q+q\alpha_p}\theta^{\mu[p]}\theta^{\nu[q]}[\eta_{\nu[q]},\omega_{\mu[p]}]\\
&=\tfrac{1}{p!q!}(-1)^{1+\alpha_p\alpha_q+q\alpha_p+p(q+\alpha_q)}[\theta^{\nu[q]}\eta_{\nu[q]},\theta^{\mu[p]}\omega_{\mu[p]}]\\
&=(-1)^{1+N_pN_q}[\eta_q,\omega_p]\;,
\end{split}    
\end{equation}
where we used the shorthand notation $\theta^{\mu[p]}:=\theta^{\mu_1}...\theta^{\mu_p}$ and $\omega_{\mu[p]}:=\omega_{[\mu_1...\mu_p]}\,$.
Both differentials $d$ and $\cD$ (and thus $\del$) obey a graded Leibniz rule w.r.t. the $N$ degree:
\begin{equation}
\del\,[\omega_p,\eta_q]=[\del\omega_p,\eta_q]+(-1)^{N_p}[\omega_p,\del\eta_q] \;,   
\end{equation}
that rules the graded Jacobi identity as well:
\begin{equation}
[[\omega_p,\omega_q],\omega_r]+(-1)^{N_p(N_q+N_r)}[[\omega_q,\omega_r],\omega_p]+(-1)^{N_r(N_p+N_q)}[[\omega_r,\omega_p],\omega_q]=0\;,    
\end{equation}
as can be seen from the definition \eqref{totalbracket} and \eqref{thetology}.
The advantage of having a dgLa structure according to the $N$ degree is apparent in applications to the tensor hierarchy: Recalling the degree assignments above, one can see that every $p-$form gauge field has $N$ degree zero, while every curvature has $N$ degree $-1\,$, all gauge parameters have $N$ degree $+1$ and so on. Hence, it is now meaningful to define a string field-like generating function of homogeneous $N$ degree zero for all the gauge fields:
\begin{equation}\label{generating A}
\cA(x,\theta) =\sum_{p=1}^\infty\tfrac{1}{p!}\,\theta^{\mu_1}...\theta^{\mu_p}\,A_{\mu_1...\mu_p}(x) \;,  
\end{equation}
as well as the corresponding degree $-1$ generating function for the curvatures $\cF(x,\theta)\,$, a degree $+1$ $\Lambda(x,\theta)$ for gauge parameters, degree $+2$ $\Xi(x,\theta)$ for trivial parameters etc.
The bi-complex \eqref{DIAGRAM} is now grouped along the diagonals in terms of the generating functions, yielding
\begin{equation}
...\xrightarrow{\del} Z_2(\Xi)\xrightarrow{\del}Z_1(\Lambda)\xrightarrow{\del}Z_0(\cA)\xrightarrow{\del}Z_{-1}(\cF)\xrightarrow{\del}\,Z_{-2}({\rm Bianchi})\xrightarrow{\del}...    
\end{equation}
where the single subscript denotes the $N$ degree. In the following we will sometimes refer to the $N$ degree as the total degree and to the original degree on $X$ as the internal degree.

\subsection{Bianchi identities}

Having identified the dgLa structure on the total space, it is now possible to prove the Bianchi identities to all orders. The degree zero field $\cA$ contains all the gauge $p-$forms:
\begin{equation}
\cA=\sum_{p=1}^\infty A_p  =\sum_{p=1}^\infty\tfrac{1}{p!}\,\theta^{\mu_1}...\theta^{\mu_p}\,A_{\mu_1...\mu_p}\;,  
\end{equation}
while all the curvatures are contained in the degree $-1$ field
\begin{equation}
\cF=\sum_{p=2}^\infty F_p=\sum_{p=2}^\infty\tfrac{1}{p!}\,\theta^{\mu_1}...\theta^{\mu_p}\,F_{\mu_1...\mu_p}\;.    
\end{equation}
Since the natural differential on the total space is the total differential  $\del\,$, it is convenient to define a shifted generating function:
\begin{equation}\label{OmegaMCdefined}
\Omega:=\cF+\cD A_1=\cD A_1+\sum_{p=2}^\infty F_p\;,    
\end{equation}
that is not a curvature, but rather a Maurer-Cartan (MC) like connection, as it will be clear by the form of the corresponding ``Bianchi identities''. The Maurer-Cartan generating fuction $\Omega$ can be defined by
\begin{equation}
\Omega=\sum_{N=1}^\infty\Omega_N=\sum_{N=1}^\infty\frac{(\scaleobj{1.5}{\iota}_\cA)^{N-1}}{N!}\,\del\cA=\del\cA+\tfrac12[\del\cA,\cA]+...   \;,\quad \scaleobj{1.5}{\iota}_\cA x:=[x,\cA] \;.
\end{equation}
Differentiating the $N-$field term gives
\begin{equation}\label{domegaN}
\del\Omega_N=-\tfrac{1}{N!}\,\sum_{k=0}^{N-2}(\scaleobj{1.5}{\iota}_\cA)^{N-2-k}\scaleobj{1.5}{\iota}_{\del\cA}(\scaleobj{1.5}{\iota}_\cA)^k\,\del\cA \;.   
\end{equation}
The first terms can be rearranged by using the graded Jacobi identity:
\begin{equation}\label{firstdomega}
\begin{split}
\del\Omega_2&=-\tfrac12\,[\del\cA,\del\cA]\;,\\
\del\Omega_3&=-\tfrac16\,\Big\{[[\del\cA,\del\cA],\cA]+[[\del\cA,\cA],\del\cA]\Big\}=-\tfrac12\,[[\del\cA,\cA],\del\cA]\;,\\
\del\Omega_4&=-\tfrac{1}{24}\,\Big\{[[[\del\cA,\del\cA],\cA],\cA]+[[[\del\cA,\cA],\del\cA],\cA]+[[[\del\cA,\cA],\cA],\del\cA]\Big\}\\
&=-\tfrac18\,[[\del\cA,\cA],[\del\cA,\cA]]-\tfrac16\,[[[\del\cA,\cA],\cA],\del\cA]
\end{split}    
\end{equation}
suggesting the general relation
\begin{equation}\label{MCrelation}
\sum_{k=0}^{N-2}(\scaleobj{1.5}{\iota}_\cA)^{N-2-k}\scaleobj{1.5}{\iota}_{\del\cA}(\scaleobj{1.5}{\iota}_\cA)^k\,\del\cA=\sum_{k=0}^{N-2}\tfrac12\,\scaleobj{0.6}{\binom{N}{k+1}}\,\left[(\scaleobj{1.5}{\iota}_\cA)^k\del\cA,(\scaleobj{1.5}{\iota}_\cA)^{N-2-k}\del\cA\right] \;,   
\end{equation}
that can be proven by induction: The first values of $N$ are given in \eqref{firstdomega} ; supposing that \eqref{MCrelation} holds for $N$ we can write for $N+1$
\begin{equation}
\begin{split}
&\sum_{k=0}^{N-1}(\scaleobj{1.5}{\iota}_\cA)^{N-1-k}\scaleobj{1.5}{\iota}_{\del\cA}(\scaleobj{1.5}{\iota}_\cA)^k\,\del\cA=[(\scaleobj{1.5}{\iota}_\cA)^{N-1},\del\cA]+\scaleobj{1.5}{\iota}_\cA\sum_{k=0}^{N-2}(\scaleobj{1.5}{\iota}_\cA)^{N-2-k}\scaleobj{1.5}{\iota}_{\del\cA}(\scaleobj{1.5}{\iota}_\cA)^k\,\del\cA\\
&=[(\scaleobj{1.5}{\iota}_\cA)^{N-1},\del\cA]+\sum_{k=0}^{N-2}\frac12\,\scaleobj{0.6}{\binom{N}{k+1}}\,\left[\left[(\scaleobj{1.5}{\iota}_\cA)^k\del\cA,(\scaleobj{1.5}{\iota}_\cA)^{N-2-k}\del\cA\right],\cA\right]\;.
\end{split}    
\end{equation}
By using the graded Jacobi identity and recalling that, by definition of $\scaleobj{1.5}{\iota}_\cA\,$, one has
\begin{equation}
[(\scaleobj{1.5}{\iota}_\cA)^kx,\cA]=(\scaleobj{1.5}{\iota}_\cA)^{k+1}x\;,    
\end{equation}
we can write
\begin{equation}
\begin{split}
&\sum_{k=0}^{N-1}(\scaleobj{1.5}{\iota}_\cA)^{N-1-k}\scaleobj{1.5}{\iota}_{\del\cA}(\scaleobj{1.5}{\iota}_\cA)^k\,\del\cA=[(\scaleobj{1.5}{\iota}_\cA)^{N-1}\del\cA,\del\cA]\\
&+\sum_{k=0}^{N-2}\tfrac12\,\scaleobj{0.6}{\binom{N}{k+1}}\Big\{\left[(\scaleobj{1.5}{\iota}_\cA)^{k+1}\del\cA, (\scaleobj{1.5}{\iota}_\cA)^{N-2-k}\del\cA\right]+\left[(\scaleobj{1.5}{\iota}_\cA)^{N-1-k}\del\cA, (\scaleobj{1.5}{\iota}_\cA)^{k}\del\cA\right]\Big\}\\
&=\sum_{k=0}^{N-1}\frac12\,\left\{\scaleobj{0.6}{\binom{N}{k+1}}+\scaleobj{0.6}{\binom{N}{k}}\right\}\,\left[(\scaleobj{1.5}{\iota}_\cA)^{k}\del\cA, (\scaleobj{1.5}{\iota}_\cA)^{N-1-k}\del\cA\right]\\
&=\sum_{k=0}^{N-1}\tfrac12\,\scaleobj{0.6}{\binom{N+1}{k+1}}\,\left[(\scaleobj{1.5}{\iota}_\cA)^{k}\del\cA, (\scaleobj{1.5}{\iota}_\cA)^{N-1-k}\del\cA\right]\;,
\end{split}
\end{equation}
thus proving \eqref{MCrelation} by induction. Substituting \eqref{MCrelation} in \eqref{domegaN} and summing over $N$ one finally determines
\begin{equation}\label{BianchiMC}
\del\Omega+\tfrac12\,[\Omega,\Omega]=0\;,    
\end{equation}
that in fact is the analog of the zero curvature condition for a Maurer-Cartan connection, rather than a Bianchi identity. In fact, thanks to the identity
\begin{equation}
e^{-X}d\,e^X=\sum_{N=1}^\infty\frac{(\scaleobj{1.5}{\iota}_X)^{N-1}}{N!}\,dX\;,    \quad \scaleobj{1.5}{\iota}_XY:=[Y,X]\;,
\end{equation}
valid for a Lie algebra valued\footnote{$\cA$ has total degree zero, so it effectively shares the properties of a Lie algebra element as far as sign factors are concerned.} field $X$ and differential $d\,$, the Maurer-Cartan string field can be rewritten in the suggestive form
\begin{equation}
\Omega=e^{-\cA}\del\,e^{\cA}\;.    
\end{equation}
In order to recover the Bianchi identities for $\cF\,$, we extract the curvature as $\Omega=\cF+\cD A_1$ and open the differential $\del=d+\cD\,$:
\begin{equation}\label{BianchiF}
\begin{split}
\del\Omega+\tfrac12\,[\Omega,\Omega]&=(d+\cD)\cF+d\cD A_1+\tfrac12\,[\cF,\cF]+[\cD A_1,\cF]+\tfrac12\,\cD[A_1,\cD A_1]  \\
&=D\cF+\cD(\cF-F_2)+\tfrac12\,[\cF,\cF]=0\;,
\end{split}    
\end{equation}
where $D\cF:=d\cF+[\cD A_1,\cF]$ and we recall that $F_2=dA_1+\tfrac12\,[\cD A_1,A_1]+\cD A_2\,$.
In fact, in the present formalism the covariant derivative (with respect to $A_1$) can be defined on any element of the total space by
\begin{equation}
Dx:=dx+[\cD A_1,x]   \;. 
\end{equation}
Thanks to the graded Jacobi identity $D$ is a degree $-1$ differential, \emph{i.e.}
\begin{equation}
D[x,y]=[Dx,y]+(-1)^{|x|}[x,Dy]\;,    
\end{equation}
and obeys
\begin{equation}
D^2x=-[\cD F_2,x]   \;. 
\end{equation}
The seemingly odd shift of $F_2$ in \eqref{BianchiF} is due to the fact that $\cD F_2$ does not actually appear in the Bianchi identities. The component form of \eqref{BianchiF} yields indeed the usual result:
\begin{equation}\label{Bianchicomponent}
DF_p+\tfrac12\,\sum_{k=2}^{p-1}[F_k,F_{p+1-k}]+\cD F_{p+1}=0 \;,\quad p\geq2\;.   
\end{equation}
Let us notice that the curvatures defined by $\cF=e^{-\cA}\,\del\,e^\cA-\cD A_1$ do not have the standard form ${F_{p+1}=DA_p+\cD A_{p+1}+...}$ that was given, for instance, in \cite{Bonezzi:2019ygf}. This is due to the democratic treatment of all gauge $p-$forms, where the vector $A_1$ does not have a special status in $\cA$ (except for the explicit shift $\Omega=\cD A_1+\cF$). The map between the two formulations is a field redefinition of all the higher gauge $p-$forms that we will give to all orders in appendix \ref{BCH section}.

\subsection{Gauge symmetry and covariant curvatures}

We are now ready to tackle the problem of finding the explicit form of the gauge transformations of $\cA$ that lead to covariant transformations for $\cF\,$. In practical applications to the tensor hierarchy, it is customary to use, for higher $p-$form gauge fields, the so called covariant variations $\Delta A_p$ rather than the actual variations $\delta A_p\,$. The covariant gauge transformations $\Delta_\lambda A_p$ are much simpler than $\delta_\lambda A_p$ and are defined, for a given $p-$form, in terms of the variations of lower forms. For instance, for the covariant variation of the two-form one has $\Delta A_2=\delta A_2+\frac12\,[\delta A_1,A_1]\,$. 
In the present formalism, the advantage of using the covariant variation  $\Delta\cA$ (whose form will be determined to all orders in $\cA$) is that a general variation of the Maurer-Cartan string field $\Omega$ has a remarkably simple form in terms of $\Delta\cA\,$:
\begin{equation}\label{deltaOmega}
\delta\Omega=\del\Delta\cA+[\Omega,\Delta\cA] \;,   
\end{equation}
as we will prove below. The transformation law \eqref{deltaOmega} allows one to determine the covariant gauge transformation $\Delta_\Lambda\cA$ in a straightforward way.

In order to prove \eqref{deltaOmega} and find the form of $\Delta\cA\,$, we take the variation of the order $N$ part $\Omega_N=\frac{1}{N!}(\scaleobj{1.5}{\iota}_\cA)^{N-1}\del\cA$:
\begin{equation}
\delta\Omega_N=\tfrac{1}{N!}\,\Big((\scaleobj{1.5}{\iota}_\cA)^{N-1}\del\delta\cA+\sum_{k=0}^{N-2}(\scaleobj{1.5}{\iota}_\cA)^{N-2-k}\scaleobj{1.5}{\iota}_{\delta\cA}(\scaleobj{1.5}{\iota}_\cA)^k\del\cA\Big)    \;.
\end{equation}
Next, we pull out a total derivative from the first term in order to get rid of $\del\delta\cA\,$:
\begin{equation}\label{deltaOmegaN}
\begin{split}
\delta\Omega_N&=\tfrac{1}{N!}\,\del\big\{(\scaleobj{1.5}{\iota}_\cA)^{N-1}\delta\cA\big\}\\
&+\tfrac{1}{N!}\,\sum_{k=0}^{N-2}\Big((\scaleobj{1.5}{\iota}_\cA)^{N-2-k}\scaleobj{1.5}{\iota}_{\delta\cA}(\scaleobj{1.5}{\iota}_\cA)^k\del\cA-(\scaleobj{1.5}{\iota}_\cA)^{N-2-k}\scaleobj{1.5}{\iota}_{\del\cA}(\scaleobj{1.5}{\iota}_\cA)^k\delta\cA\Big)\;.    
\end{split}    
\end{equation}
The total $\del$ term suggests the ansatz for the covariant variation:
\begin{equation}\label{DeltaA}
\Delta\cA:=\sum_{N=1}^\infty\tfrac{1}{N!}\,(\scaleobj{1.5}{\iota}_{\cA})^{N-1}\delta\cA=e^{-\cA}\delta\,e^{\cA}=\delta\cA+\tfrac12\,[\delta\cA, \cA]+...    
\end{equation}
In order to prove \eqref{deltaOmega}, the second line of \eqref{deltaOmegaN} has to be equal to
\begin{equation}
[\Omega,\Delta\cA]\rvert_N=\sum_{k=0}^{N-2}\tfrac{1}{(k+1)!(N-1-k)!}\,\left[(\scaleobj{1.5}{\iota}_{\cA})^k\del\cA, (\scaleobj{1.5}{\iota}_{\cA})^{N-2-k}\delta\cA\right]  \;.  
\end{equation}
For this, we are going to prove by induction the identity
\begin{equation}\label{deltaOmegainduc}
\begin{split}
&\sum_{k=0}^{N-2}\Big\{(\scaleobj{1.5}{\iota}_\cA)^{N-2-k}\scaleobj{1.5}{\iota}_{\delta\cA}(\scaleobj{1.5}{\iota}_\cA)^k\del\cA-(\scaleobj{1.5}{\iota}_\cA)^{N-2-k}\scaleobj{1.5}{\iota}_{\del\cA}(\scaleobj{1.5}{\iota}_\cA)^k\delta\cA\Big\}\\
&=\sum_{k=0}^{N-2}\scaleobj{0.6}{\binom{N}{k+1}}\,\left[(\scaleobj{1.5}{\iota}_{\cA})^k\del\cA, (\scaleobj{1.5}{\iota}_{\cA})^{N-2-k}\delta\cA\right]\;.
\end{split}    
\end{equation}
The lowest $N$ cases give
\begin{equation}
\begin{split}
N=2\qquad & [\del\cA, \delta\cA]-[\delta\cA, \del\cA]=2\,[\del\cA,\delta\cA]\;,\\
N=3\qquad & [[\del\cA,\delta\cA],\cA]-[[\delta\cA,\del\cA],\cA]+[[\del\cA,\cA],\delta\cA]-[[\delta\cA,\cA],\del\cA]\\
&=2\,[[\del\cA,\delta\cA],\cA]+[[\del\cA,\cA],\delta\cA]+[\del\cA,[\delta\cA,\cA]]\\
&=3\,[[\del\cA,\cA],\delta\cA]+3\,[\del\cA,[\delta\cA,\cA]]
\end{split}    
\end{equation}
by using graded anti-symmetry and Jacobi. Supposing that \eqref{deltaOmegainduc} holds for $N\,$, for $N+1$ we obtain
\begin{equation}
\begin{split}
&\sum_{k=0}^{N-1}\Big\{(\scaleobj{1.5}{\iota}_\cA)^{N-1-k}\scaleobj{1.5}{\iota}_{\delta\cA}(\scaleobj{1.5}{\iota}_\cA)^k\del\cA-(\scaleobj{1.5}{\iota}_\cA)^{N-1-k}\scaleobj{1.5}{\iota}_{\del\cA}(\scaleobj{1.5}{\iota}_\cA)^k\delta\cA\Big\}\\
&=\left[(\scaleobj{1.5}{\iota}_\cA)^{N-1}\del\cA,\delta\cA\right]-\left[(\scaleobj{1.5}{\iota}_\cA)^{N-1}\delta\cA,\del\cA\right] \\
&+\scaleobj{1.5}{\iota}_\cA\sum_{k=0}^{N-2}\Big\{(\scaleobj{1.5}{\iota}_\cA)^{N-2-k}\scaleobj{1.5}{\iota}_{\delta\cA}(\scaleobj{1.5}{\iota}_\cA)^k\del\cA-(\scaleobj{1.5}{\iota}_\cA)^{N-2-k}\scaleobj{1.5}{\iota}_{\del\cA}(\scaleobj{1.5}{\iota}_\cA)^k\delta\cA\Big\}\\
&=\left[(\scaleobj{1.5}{\iota}_\cA)^{N-1}\del\cA,\delta\cA\right]-\left[(\scaleobj{1.5}{\iota}_\cA)^{N-1}\delta\cA,\del\cA\right]+\scaleobj{1.5}{\iota}_\cA\sum_{k=0}^{N-2}\scaleobj{0.6}{\binom{N}{k+1}}\,\left[(\scaleobj{1.5}{\iota}_{\cA})^k\del\cA, (\scaleobj{1.5}{\iota}_{\cA})^{N-2-k}\delta\cA\right]\;,
\end{split}    
\end{equation}
and, by using graded Jacobi and the first two terms for the boundaries of the sum, it finally yields
\begin{equation}
\begin{split}
&\sum_{k=0}^{N-1}\Big\{(\scaleobj{1.5}{\iota}_\cA)^{N-1-k}\scaleobj{1.5}{\iota}_{\delta\cA}(\scaleobj{1.5}{\iota}_\cA)^k\del\cA-(\scaleobj{1.5}{\iota}_\cA)^{N-1-k}\scaleobj{1.5}{\iota}_{\del\cA}(\scaleobj{1.5}{\iota}_\cA)^k\delta\cA\Big\}\\
&=\sum_{k=0}^{N-1}\left\{\scaleobj{0.6}{\binom{N}{k+1}}+\scaleobj{0.6}{\binom{N}{k}}\right\}\,\left[(\scaleobj{1.5}{\iota}_{\cA})^k\del\cA, (\scaleobj{1.5}{\iota}_{\cA})^{N-1-k}\delta\cA\right]\\
&=\sum_{k=0}^{N-1}\left\{\scaleobj{0.6}{\binom{N+1}{k+1}}\right\}\,\left[(\scaleobj{1.5}{\iota}_{\cA})^k\del\cA, (\scaleobj{1.5}{\iota}_{\cA})^{N-1-k}\delta\cA\right]\;,
\end{split}    
\end{equation}
thus proving \eqref{deltaOmega} with $\Delta\cA$ given by \eqref{DeltaA}.

At this point we introduce the degree $+1$ gauge parameter string field
\begin{equation}
\Lambda=\sum_{p=0}^\infty \lambda_p=\sum_{p=0}^\infty\tfrac{1}{p!}\,\theta^{\mu_1}...\theta^{\mu_p}\,\lambda_{\mu_1...\mu_p}\;, 
\end{equation}
and determine the gauge transformation $\Delta_\Lambda\cA\,$. To do so, let us notice that the variation \eqref{deltaOmega} defines a new differential:
\begin{equation}
\delta\Omega=\del_\Omega\Delta\cA\;,\quad \del_\Omega\, x:=\del x+[\Omega,x]    
\end{equation}
that squares to zero thanks to the Bianchi identity (zero curvature relation) \eqref{BianchiMC}:
\begin{equation}
\del_\Omega^2=\del\Omega+\tfrac12\,[\Omega,\Omega]=0\;. \end{equation}
If we took $\Delta_\Lambda\cA=\del_\Omega\Lambda\,$, then $\Omega$ would be completely invariant. However, this is not quite the right transformation, since in the term $\del\Lambda$  there is a zero-form $\cD\lambda_0$ that does not correspond to the variation of any $A_p\,$. The correct gauge transformation is thus given by
\begin{equation}\label{gaugetransA}
\Delta_\Lambda\cA=\del_\Omega\Lambda-\cD\lambda_0=\del\Lambda+[\Omega,\Lambda]-\cD\lambda_0  \;.  
\end{equation}
This shift is ultimately responsible for the curvatures to transform only w.r.t. the $\lambda_0$ parameter. Indeed, by using $\del_\Omega^2=0$  and \eqref{gaugetransA} one obtains
\begin{equation}
\delta_\Lambda\Omega=[\cD\lambda_0,\Omega]-d\cD\lambda_0\end{equation}
that, rewritten in terms of the curvature $\cF=\Omega-\cD A_1\,$, takes the familiar form
\begin{equation}
\delta_\Lambda\cF=[\cD\lambda_0,\cF]\equiv\cL_{\lambda_0}\cF    
\end{equation}
upon using
\begin{equation}
\delta_\Lambda A_1\equiv \Delta_\Lambda A_1=D\lambda_0+\cD\lambda_1    \;, 
\end{equation}
and recalling that $Dx=dx+[\cD A_1,x]\,$.
The covariant gauge transformation \eqref{gaugetransA} can also be rewritten in a more familiar form by using $\Omega=\cF+\cD A_1$ and $\del=d+\cD\,$:
\begin{equation}
\Delta_\Lambda\cA=D\Lambda+[\cF,\Lambda]+\cD(\Lambda-\lambda_0) \;,   
\end{equation}
that in components has the usual form
\begin{equation}
\Delta_\Lambda A_p=D\lambda_{p-1}+\sum_{k=0}^{p-2}\left[\lambda_k, F_{p-k}\right]+\cD\lambda_p\;,\quad p\geq1\;.
\end{equation}
The reducibility of the gauge symmetries is also manifest, since from nilpotency of $\del_\Omega$ it is obvious that a gauge parameter of the form
$\Lambda=\del_\Omega\Xi$ is trivial\footnote{Notice that the lowest order part of such a $\Lambda$ is $\lambda_0=\cD\xi_0\,$, thus making the $\lambda_0$ shift in \eqref{gaugetransA} trivial as well.}: $\Delta_{\del_\Omega\Xi}\cA=0\,$, and the chain of reducibility continues indefinitely. 

To conclude this section, let us notice that when evaluating the above expressions one encounters terms of the form $\cD\cD u\,$, with $u\in X_1\,$, typically with $u=A_1, \lambda_0,..\,$, and these terms are not well defined, since they involve $\cD$ acting on the lowest space $X_0\,$. However, such terms only arise as the image of $\cD^2\,$, and it is sufficient to extend nilpotency of $\cD$ by declaring $\cD^2 u=0$ for $u\in X_1\,$, without introducing non trivial spaces in negative degree. This will change in the next section, where the issue will be addressed.

\subsection{Inclusion of scalars}

By looking at the bi-complex diagram \eqref{DIAGRAM} one can see that something is missing. Indeed, the right boundary stops with the spaces $Z^1_{[p]}\,$, meaning that we have introduced the space $X_0$ but there are no fields taking values in it. Moreover, from the point of view of the string field $\cA(x,\theta)$ there is no natural reason for it to start with the one-form valued in $X_1$: $\cA=\theta^\mu A_\mu+...\,$, given  that the space $X_0=\frak g$ is now available. The most natural attempt to describe the scalar geometry appears thus to repeat the aforementioned construction by letting the string field have an arbitrary expansion in powers of $\theta^\mu\,$. We shall thus define
\begin{equation}
\cA^{\phi}(x,\theta):=\sum_{p=0}^\infty\tfrac{1}{p!}\,\theta^{\mu_1}...\theta^{\mu_p}\,A_{\mu_1...\mu_p}(x)=\phi(x)+\cA(x,\theta)\;,    
\end{equation}
where we identified $\phi\equiv A_0$ taking values in the Lie algebra $\frak g\,$. Correspondingly, the Maurer-Cartan string field gets modified to
\begin{equation}
\Omega^\phi:=\sum_{N=1}^\infty\frac{(\scaleobj{1.5}{\iota}_{\cA^\phi})^{N-1}}{N!}\del\cA^\phi  =e^{-\cA^\phi}\del\,e^{\cA^\phi} \;, 
\end{equation}
and obeys the same zero curvature condition:
\begin{equation}\label{MCOmegaphi}
\del\Omega^\phi+\tfrac12\,[\Omega^\phi,\Omega^\phi]=0 \;,
\end{equation}
since the proof in the previous section does not depend on the form degrees  of $\cA$ and $\Omega\,$. 

In order to make sense of the above expressions, one has to define the action of $\cD$ on the Lie algebra $X_0\,$. Up to this point, the only occurrence of $\cD$ acting on $X_0$ was of the form $\cD\cD u\,$, with $u\in X_1\,$, as we discussed above. We will discuss in detail the extension of $\cD$ to $X_0\,$, but for the moment we just notice that the Maurer-Cartan form $\Omega^\phi$ acquires a zero-form component $\Omega^\phi_0=e^{-\phi}\cD\,e^\phi\,$, that takes values in a new space $X_{-1}$ to be defined in the following. 

Before doing so, we will address the issue of the field basis $\cA^\phi\,$: indeed, this construction does not lead directly to the standard description of the scalar manifold for two reasons:
\begin{itemize}
\item[i)] The gauge covariant curvatures are not expressed anymore by $\cF=\Omega^{\phi}-\cD A_1\,$, and
\item[ii)] The field basis for the gauge $p-$forms is different from the standard one.
\end{itemize}
To begin with, we shall evaluate the one-form component of $\Omega^\phi$ that, compared to the previous case $\Omega_1=\cD A_1\,$, now receives infinitely many contributions in $\phi\,$:
\begin{equation}
\Omega^\phi_1=\sum_{N=1}^\infty\frac{(\scaleobj{1.5}{\iota}_\phi)^{N-1}}{N!}(d\phi+\cD A_1) =e^{-\phi}d\,e^\phi+   \sum_{N=1}^\infty\frac{(\scaleobj{1.5}{\iota}_\phi)^{N-1}}{N!}\cD A_1\;.
\end{equation}
By defining the group valued scalar field
\begin{equation}
\cV:=e^\phi\;\in\,G  \;,  
\end{equation}
the first term is the familiar Maurer-Cartan one-form on the group manifold $G\,$: $\cV^{-1}d\cV\,$. The second term, however, cannot be expressed in terms of $\cV$ and therefore cannot be interpreted as a gauge covariant improvement of the former. 

To remedy this, we shall redefine the one-form $A_1$ to all orders in $\phi\,$, as it can be inferred by rewriting the first few terms in the series:
\begin{equation}
\begin{split}
\sum_{N=1}^\infty\frac{(\scaleobj{1.5}{\iota}_\phi)^{N-1}}{N!}\cD A_1 &= \cD A_1+\tfrac12\,[\cD A_1,\phi]+\tfrac16\,[[\cD A_1,\phi],\phi]+\cO(\phi^3)  \\
&=\cD\left\{A_1-\tfrac12\,[A_1,\phi]+\tfrac16\,[[A_1,\phi],\phi]\right\}+[\cD(A_1-\tfrac12\,[A_1,\phi]),\phi]\\
&\quad+\tfrac12\,[[\cD A_1,\phi],\phi]+\cO(\phi^3)\;,
\end{split}    
\end{equation}
suggesting that the field redefinition
\begin{equation}\label{A'firstorders}
A_1'=A_1-\tfrac12\,[A_1,\phi]+\tfrac16\,[[A_1,\phi],\phi]+\cO(\phi^3)    
\end{equation}
will bring the above expression to
\begin{equation}\label{tentativescalarP}
\sum_{N=1}^\infty\frac{(\scaleobj{1.5}{\iota}_\phi)^{N-1}}{N!}\cD A_1=\sum_{N=0}^\infty\frac{(\scaleobj{1.5}{\iota}_\phi)^{N}}{N!}\cD A'_1=e^{-\phi}\cD A'_1\,e^\phi\;,
\end{equation}
where the last identification is allowed by the Lie algebra identity
\begin{equation}\label{BCHY}
e^{-X}Y\,e^X=\sum_{N=0}^\infty\frac{(\scaleobj{1.5}{\iota}_X)^N}{N!}Y=e^{\scaleobj{1.5}{\iota}_X}Y \;,\quad \scaleobj{1.5}{\iota}_XY=[Y,X]\;.   
\end{equation}
We will find in the following the field redefinition to all orders such that \eqref{tentativescalarP} holds. For now, assuming \eqref{tentativescalarP}, we notice that $\Omega_1^\phi$ takes the form of a gauge covariantized Maurer-Cartan one-form on the group manifold $G\,$:
\begin{equation}\label{MCOmega1}
\Omega^\phi_1=\cV^{-1}(d+\cD A'_1)\cV\;.    
\end{equation}
In order to proceed further and identify the gauge covariant curvatures, let us examine the integrability condition of $\Omega_1^\phi\,$.
Extracting the two-form component from \eqref{MCOmegaphi} one has
\begin{equation}
d\Omega_1^\phi+\tfrac12\,[\Omega_1^\phi,\Omega_1^\phi]+\cD\Omega_2^\phi=0\;.    
\end{equation}
For this integrability condition to be defined in terms of the group manifold scalar $\cV\,$, one should show that the two-form $\Omega_2^\phi$ can be entirely expressed (in the primed basis) in terms of the gauge covariant curvature $F_2$ and $\cV$ itself. Again, this is possible (as it will be shown to all orders) by means of a field redefinition, whose first orders can be determined by working on $\Omega_2^\phi$ at cubic order in the fields:
\begin{equation}
\begin{split}
\Omega_2^\phi&=dA_1+\cD A_2+\tfrac12\,[dA_1+\cD A_2,\phi]+\tfrac12\,[\cD A_1,A_1]+\tfrac12\,[d\phi,A_1]\\
&\quad+\tfrac16\,[[dA_1+\cD A_2,\phi],\phi]+\tfrac16\,[[d\phi,\phi],A_1]+\tfrac16\,[[d\phi,A_1],\phi]\\
&\quad+\tfrac16\,[[\cD A_1,A_1],\phi]+\tfrac16\,[[\cD A_1,\phi],A_1]+\cO({\rm field}^4)\\
&=F'_2+[F'_2,\phi]+\tfrac12\,[[F'_2,\phi],\phi]+\cO(\phi^3)\;,
\end{split}    
\end{equation}
where the prime on $F_2$ means that it is written in terms of primed gauge fields:
\begin{equation}
F'_2=dA'_1+\tfrac12\,[\cD A'_1,A'_1]+\cD A'_2  \;,  
\end{equation}
with the field redefinition determined up to cubic order:
\begin{equation}\label{fieldredefcubic}
\begin{split}
A'_1&=A_1-\tfrac12\,[A_1,\phi]+\tfrac16\,[[A_1,\phi],\phi]+\cO({\rm field}^4)\;,\\
A'_2&=A_2-\tfrac12\,[A_2,\phi]+\tfrac16\,[[A_2,\phi],\phi]-\tfrac{1}{12}\,[[\phi,A_1],A_1]+\cO({\rm field}^4)\;.
\end{split}    
\end{equation}
In terms of the primed $A_1$ and $A_2$ one should obtain 
\begin{equation}
\Omega_2^\phi=\sum_{N=0}^\infty\frac{(\scaleobj{1.5}{\iota}_\phi)^N}{N!}F'_2=e^{-\phi}F'_2\,e^\phi  \;,  
\end{equation}
yielding the gauged version of the integrability condition for the MC form $\Omega_1^\phi\,$:
\begin{equation}
d\Omega_1^\phi+\tfrac12\,[\Omega_1^\phi,\Omega_1^\phi]+\cV^{-1}\cD F'_2\,\cV=0  \;.  
\end{equation}
Before giving the all order form for the field redefinition and the above statements, let us notice that the scalar manifold parametrized by $\phi$ or $\cV$ is a group manifold, since $\phi$ takes values in the Lie algebra $\frak g$ and $\cV=e^\phi$ in the Lie group $G\,$. In actual applications, however, the scalar manifold is rather a coset manifold $G/H\,$. In order to describe such a structure, we need extra information: in particular, one needs to specify the subalgebra $\frak h \subset \frak g$ that plays the role of the isotropy algebra of the coset space. Once this is given, it is possible to define the so called $H-$connection $Q$ by projecting the MC form $\Omega_1^\phi$ to $\frak h\,$:
\begin{equation}
Q:=\left[\cV^{-1}(d+\cD A'_1)\cV\right]\rvert_{\frak h}\;, 
\end{equation}
while its complement defines the pullback $P$ of the vielbein of the coset manifold $G/H\,$:
\begin{equation}
P:=\cV^{-1}(d+\cD A'_1)\cV-Q    \;,
\end{equation}
in the standard form for gauged coset non-linear sigma models.

In order to find the field redefinition $\cA\to\cA'(\cA,\phi)$ to all orders, let us invert the cubic expression \eqref{fieldredefcubic}:
\begin{equation}
\begin{split}
A_1&=A'_1+\tfrac12\,[A'_1,\phi]+\tfrac{1}{12}\,[[A'_1,\phi],\phi]+\cO({\rm field}^4)\;,\\
A_2&=A'_2+\tfrac12\,[A'_2,\phi]+\tfrac{1}{12}\,[[A'_2,\phi],\phi]+\tfrac{1}{12}\,[[\phi,A'_1],A'_1]+\cO({\rm field}^4)\;.    
\end{split}    
\end{equation}
These are precisely the first terms in the Baker-Campbell-Hausdorff series, that are produced by the all order field redefinition
\begin{equation}\label{fieldredefphi}
\cA=\ln\left(e^{\cA'}e^\phi\right)-\phi=\cA'+\tfrac12\,[\cA',\phi]+\tfrac{1}{12}\,[[\cA',\phi],\phi]+\tfrac{1}{12}\,[[\phi,\cA'],\cA']+...   
\end{equation}
Indeed, by using 
\begin{equation}
e^{-Y}e^{-X}d(e^X e^Y) =e^{-Y}d\,e^Y+e^{-Y}\left(e^{-X}d\,e^X\right)e^Y   
\end{equation}
for Lie algebra valued fields $X$ and $Y\,$, one can derive
\begin{equation}
\begin{split}
\Omega^\phi&=e^{-\cA^\phi}\del\,e^{\cA^\phi}=e^{-\cA-\phi}\del\,e^{\cA+\phi} =e^{-\phi}e^{-\cA'}\del\left(e^{\cA'}e^\phi\right)  \\
&= e^{-\phi}\del\,e^\phi+e^{-\phi}\left(e^{-\cA'}\del\,e^{\cA'}\right)e^\phi=e^{-\phi}\del\,e^\phi+e^{-\phi}\Omega'e^\phi\\
&=\cV^{-1}(d+\cD A'_1)\cV+\cV^{-1}\cF'\cV+\Omega_0\;.
\end{split}    
\end{equation}
This in turn gives also the relation between $\Omega^\phi$ and the gauge covariant curvatures:
\begin{equation}
\begin{split}
\Omega^\phi_1&=\cV^{-1}(d+\cD A'_1)\cV=P+Q\;,\\
\Omega^\phi_p&=\cV^{-1}F'_p\cV\;,\quad p\geq2\;.
\end{split}    
\end{equation}
Having determined the field basis $\cA'\,$, which is the most natural for the inclusion of scalars, we shall drop the primes and write the Maurer-Cartan field $\Omega$ as
\begin{equation}\label{Omegawithphifactor}
\Omega=e^{-\phi}e^{-\cA}\,\del\left(e^\cA e^\phi\right)=\Omega_0+\cV^{-1}(d+\cD A_1)\cV+\cV^{-1}\cF\,\cV\;,   \end{equation}
where $\cF=e^{-\cA}\del e^\cA-\cD A_1\,$ and $\cV=e^\phi\in G\,$. 
We are now ready to define the extension of the differential $\cD$ to $X_0\,$, that will also allow to properly interpret the zero-form part of $\Omega\,$.

\subsection{The space $X_{-1}$ and the embedding tensor}

In order to define the action of $\cD$ on $X_0$ one has to introduce a new space $X_{-1}$ in degree $-1\,$, such that $\cD:X_0\to X_{-1}$ is well-defined. The simplest choice would be to declare that the new space only consists of the trivial vector, $X_{-1}=\{0\}\,$, yielding $\cD x=0$ for any $x\in \mathfrak{g}\,$. This choice leads to a completely consistent model that, however, is not general enough for gauged supergravity since  typically the group $G$ is only a symmetry group of the ungauged theory. If one transforms the gauge fields as $\cA\to g\,\cA\,g^{-1}$ with a constant $G$ parameter, the choice $\cD g=0$ would lead to $\Omega$ being fully $G-$invariant even in the gauged theory. This is unacceptable if one wants to describe general gauged supergravities where, after gauging a subgroup $G_0\subset G\,$, the symmetry gets reduced to (local) $G_0$ transformations. 

We shall thus introduce a new non-trivial space $X_{-1}\,$, and we will make no further assumptions about lower negative degree spaces that can be produced by recursive brackets of elements in $X_{-1}\,$, since they will never appear in the tensor hierarchy.
At this point, the differential $\cD$ itself can be used to define an element $\Theta\in X_{-1}$ satisfying $[\Theta,\Theta]=0$ via
\begin{equation}
[\Theta,u]:=\cD u\;,\quad\forall u\in X\;.    
\end{equation}
Such an element $\Theta$ always exists since this relation is integrable: acting with $\cD$ one obtains 
\be
 0 = \cD^2u = [\cD\Theta, u]-[\Theta,\cD u] = [[\Theta,\Theta],u] -\cD^2 u = [[\Theta,\Theta],u]\;, 
\ee
which is satisfied since we assume $[\Theta,\Theta]=0$.  
The new element $\Theta$ plays the role of a generalized embedding tensor. In order to make contact with the usual expressions in the literature one can introduce basis vectors for the lower spaces as follows:
\begin{equation}
\begin{array}{cccccc}
{\rm degree} & -1 & 0 & +1 & +2 & \cdots\\
{} & e_A & t_\alpha & e_M & e_I & \cdots\;,
\end{array}    
\end{equation}
and define the structure constants by
\begin{equation}
\begin{split}
&[t_\alpha,e_A]=T_{\alpha\,A}{}^B\,e_B \;,\quad [t_\alpha,t_\beta]=f_{\alpha\beta}{}^\gamma\,t_\gamma\;,\quad [t_\alpha,e_M]=T_{\alpha\,M}{}^N\,e_N\;,\quad [t_\alpha,e_I]=T_{\alpha\,I}{}^J\,e_J\;,\\  
&[e_A,e_M]=F_{A\,M}^\alpha\,t_\alpha\;,\quad [e_A,e_I]=F_{A\,I}^M\,e_M\;,\quad{\rm etc.}
\end{split}    
\end{equation}
The element $\Theta$ can be expanded as $\Theta=\Theta^A\,e_A$ and yields the usual embedding tensor when acting on the Leibniz space $X_1\,$:
\begin{equation}\label{embeddingusual}
[\Theta,u]=t_\alpha\,\theta_M{}^\alpha\,u^M\;,\quad \theta_M{}^\alpha:=\Theta^A\,F_{A\,M}^\alpha\;.    
\end{equation}
In this context, the linear (representation) constraint obeyed by $\theta_M{}^\alpha$ is included in the statement that $\Theta^A$ gives the actual $\mathfrak{g}-$representation $X_{-1}$ of the embedding tensor, while $\theta_M{}^\alpha\,$, as defined in \eqref{embeddingusual}, expresses it as a sub-representation in the tensor product $X_0\otimes X_1^*\,$. The usual quadratic constraint is provided by the Jacobi identity
\begin{equation}
[[\Theta,x],u]=[\Theta,[x,u]]-[x,[\Theta,u]]\;,\quad x\in X_0\;,u\in X_1
\end{equation}
for $x=[\Theta,v]\,$, since the left hand side then vanishes thanks to $[\Theta,\Theta]=0\,$, and the right hand side gives
\begin{equation}
u^M\,v^P\,t_\alpha\,\{\theta_P{}^\beta\,T_{\beta\,M}{}^N\,\theta_M{}^\alpha-\theta_P{}^\beta\,f_{\beta\gamma}{}^\alpha\,\theta_M{}^\gamma\} =0\;.   
\end{equation}
The action of $\Theta$ on higher spaces yields the $\cY$ tensors defined in the literature as intertwiners. For instance, on the basis elements $e_I$ of $X_2$ one has
\begin{equation}
[\Theta,e_I]=\cY_I{}^M\,e_M\;,\quad \cY_I{}^M:=\Theta^A\,F_{A\,I}^M  \;, \end{equation}
yielding relations of the usual form $\theta_M{}^\alpha\,\cY_I{}^M=0\,$, thanks to $[\Theta,\Theta]=0\,$.
Finally, acting with the differential $\cD$ on a Lie algebra element $x\in X_0$ gives the corresponding $\mathfrak{g}-$variation of the embedding tensor:
\begin{equation}
\cD x=[\Theta,x]=-[x,\Theta]=-\rho_x(\Theta)\equiv-\delta_x\Theta\;,    
\end{equation}
or, in components, ${\delta_x\Theta=e_B(x^\alpha\,T_{\alpha\,A}{}^B\,\Theta^A)}\,$.
This last relation, in particular, shows that truncating the complex by demanding $X_{-2}=\{0\}$ poses a problem: For any $x\in\mathfrak{g}$ one can define $\delta_x\Theta=\rho_x\Theta\in X_{-1}\,$, that can be exponentiated to $\Theta':=e^{\rho_x}\Theta=e^x\,\Theta\, e^{-x}\,$. While $[\Theta',\Theta']=0$ (that holds regardless of the triviality of $X_{-2}$) simply states $G-$covariance of the quadratic constraint, demanding $X_{-2}=\{0\}$ would imply that $\delta_x\Theta$ itself obeys $[\delta_x\Theta,\delta_x\Theta]=0$. 
However, in general $G-$covariance only implies that $\Theta'=\Theta+\delta_x\Theta+\cdots$ satisfies the quadratic constraint, not $\delta_x\Theta$ separately. 
For this reason, as stated above, we will not make any further assumption about triviality of lower spaces in negative degree.

It is now possible to interpret the zero-form part of the Maurer-Cartan field $\Omega\,$: one has
\begin{equation}
\Omega_0=e^{-\phi}\cD e^\phi= [\Theta,\phi]+\tfrac12\,[[\Theta,\phi],\phi]+...= e^{-\phi}\,\Theta\,e^\phi-\Theta  \;,
\end{equation}
and the first term is the dressing of the embedding tensor by $\cV\,$, that is usually named the $T-$tensor, \emph{i.e.}
\begin{equation}
T:=\cV^{-1}\Theta\cV    \;.
\end{equation}
In order to get rid of the shift by $\Theta$ it is more convenient to consider the full operator 
\begin{equation}
\del_\Omega=\del+\Omega\;,    
\end{equation}
where $\Omega$ is meant to act via the bracket, since the $\cD$ differential in $\del$ precisely cancels the $\Theta$ shift in $\Omega_0\,$, resulting in 
\begin{equation}
\del_\Omega=\del+e^{-\phi}e^{-\cA}\,\del\left(e^\cA e^\phi\right)=e^{-\phi}e^{-\cA}\,\del\,e^\cA e^\phi = d+T+\cV^{-1}(d+\cD A_1)\cV+\cV^{-1}\cF\cV   \;,
\end{equation}
where in the second expression the differential $\del$ is meant to act through.

\subsection{Coset structure, Bianchi identities and gauge symmetries}

In order to specify the coset structure $G/H$ of the scalar manifold we denote by $\mathfrak{h}$ the Lie algebra of the subgroup $H\,$, and by $\mathfrak{p}$ its complement with respect to the full $\mathfrak{g}\,$. We shall also assume that $\mathfrak{p}$ is an $\mathfrak{h}-$representation, so that in general
\begin{equation}
[\mathfrak{h},\mathfrak{h}]\subset \mathfrak{h}\;,\quad [\mathfrak{h},\mathfrak{p}]\subset \mathfrak{p}\;,\quad [\mathfrak{p},\mathfrak{p}]\subset \mathfrak{h}+\mathfrak{p} \;,
\end{equation}
while symmetric spaces are characterized by $[\mathfrak{p,\mathfrak{p}}]\subset \mathfrak{h}\,$.
The $G-$valued scalar $\cV(x)$ transforms from the left under global $G-$transformations, and from the right under local $H-$transformations:
\begin{equation}
\cV'(x)=g\,\cV(x)\,h(x)   \;. 
\end{equation}
The gauge fields contained in $\cA$ are inert under local $H-$transformations, and transform (in the ungauged limit) as $\cA'=g\cA g^{-1}$ under global $G\,$. In the ungauged theory one has $\del=d\,$, and the corresponding $\Omega$ is indeed invariant under the full duality group. In the gauged version the Maurer-Cartan field $\Omega$ (and in particular the curvature $\cF$) is not $G-$invariant, since $\cD g\neq0\,$, and transforms as
\begin{equation}
\Omega'=\Omega+\cV^{-1}e^{-\cA}(g^{-1}\,\cD g)e^\cA\cV  \;,  
\end{equation}
showing how the gauging procedure breaks the global symmetry group $G\,$.
Under the local $H$ subgroup $\Omega$ transforms as
\begin{equation}
\Omega'=h^{-1}\del h+h^{-1}\Omega\, h    \;.
\end{equation}
In particular, the zero-form part gives
\begin{equation}
\Omega_0'=h^{-1}\cV^{-1}\cD (\cV h)=h^{-1}\Omega_0 h+h^{-1}\cD h=h^{-1}\Omega_0 h+h^{-1}\Theta h-\Theta  \;,  
\end{equation}
implying $H-$covariance of the $T-$tensor:
\begin{equation}
T=\Theta+\Omega_0\;,\quad T'=h^{-1}Th    \;.
\end{equation}

The part of $\Omega_1$ that transforms with the inhomogeneous term $h^{-1}dh$ defines the composite $H-$connection one-form $Q\,$:
\begin{equation}
Q:=[\cV^{-1}(d+\cD A_1)\cV]\rvert_\mathfrak{h} \;,   
\end{equation}
while the complement one-form
\begin{equation}
P:=\cV^{-1}(d+\cD A_1)\cV-Q \in\mathfrak{p}   
\end{equation}
defines the pullback of the vielbein $P$ on the coset $G/H\,$, that transforms as a tensor under local $H-$transformations: $P'=h^{-1}Ph\,$, and $\Omega$ takes the form
\begin{equation}\label{Omegadecomposed}
\Omega=T-\Theta+P+Q+\cV^{-1}\cF\,\cV  \;,
\end{equation}
or, in terms of the operator $\del_\Omega\,$,
\begin{equation}\label{delOmegadecomposed}
\del_\Omega=D_Q+T+P+\cV^{-1}\cF\cV  \;,  
\end{equation}
where the $H-$covariant derivative is defined by $D_Q:=d+[Q,\cdot]\,$.

Since $\Omega$ is still of the form $\cG^{-1}\del\cG$ with $\cG:=e^\cA e^\phi\,$, Bianchi identities can still be extracted from the zero-curvature relation
\begin{equation}\label{Omegaflat}
\del\Omega+\tfrac12\,[\Omega,\Omega]=0    \;,
\end{equation}
or, equivalently, nilpotency of $\del_\Omega\,$.
By using the decomposition \eqref{delOmegadecomposed} one finds
\begin{equation}\label{BigBianchimisterious}
\begin{split}
0=\del_\Omega^2&=R(Q)+D_QT+D_QP+D_Q(\cV^{-1}\cF\cV)+[T,P]\\
&+\tfrac12\,[P,P]+\tfrac12\,\cV^{-1}[\cF,\cF]\cV+[P,\cV^{-1}\cF\cV]+[T,\cV^{-1}\cF\cV]   \;, 
\end{split}
\end{equation}
where we used $[T,T]=\cV^{-1}[\Theta,\Theta]\cV=0\,$, and defined the $H-$curvature by
\begin{equation}
D_Q^2=R(Q)=dQ+\tfrac12\,[Q,Q]   \;. 
\end{equation}
In order to present \eqref{BigBianchimisterious} in a more familiar form, we use $[T,\cV^{-1}\cF\cV]=\cV^{-1}[\Theta,\cF]\cV=\cV^{-1}\cD\cF\cV$ and
\begin{equation}
\begin{split}
D_Q(\cV^{-1}\cF\cV)&=\cV^{-1}d\cF\cV-[\cV^{-1}d\cV,\cV^{-1}\cF\cV]+[Q,\cV^{-1}\cF\cV]\\
&=\cV^{-1}d\cF\cV-[P,\cV^{-1}\cF\cV]+\cV^{-1}[\cD A_1,\cF]\cV\\
&=\cV^{-1}D\cF\cV-[P,\cV^{-1}\cF\cV]\;,
\end{split}    
\end{equation}
where we recall that $D=d+[\cD A_1,\cdot]$ is the gauge covariant derivative. The generalized Bianchi identity can then be recast in the form
\begin{equation}
\begin{split}
D_QT+[T,P]+R(Q)+D_QP+\tfrac12\,[P,P]+\cV^{-1}\left(D\cF+\tfrac12\,[\cF,\cF]+\cD\cF\right)\cV    =0\;.
\end{split}    
\end{equation}
Splitting the Bianchi identity in terms of form degrees we have the one-form integrability condition for the $T-$tensor:
\begin{equation}
D_QT+[T,P]=0\;,    
\end{equation}
that, given $T=\cV^{-1}\Theta\cV\,$, is equivalent to the statement of $\Theta$ being constant: $d\Theta=0\,$. The two-form part further splits along $\mathfrak{h}$ and $\mathfrak{p}\,$, yielding the $H-$curvature
\begin{equation}\label{RQ}
R(Q)+\tfrac12[P,P]\rvert_\mathfrak{h}+\big(\cV^{-1}\cD F_2\cV  \big) \rvert_\mathfrak{h}=0 
\end{equation}
and the integrability condition for $P\,$:
\begin{equation}\label{PBianchi}
D_QP+\tfrac12[P,P]\rvert_\mathfrak{p}+\big(\cV^{-1}\cD F_2\cV  \big) \rvert_\mathfrak{p}=0  \;,  
\end{equation}
as well as the usual Bianchi identities for the curvatures:
\begin{equation}\label{Bianchirepeated}
D\cF+\tfrac12[\cF,\cF]+\cD(\cF-F_2)=0\;.    
\end{equation}

Having established the Bianchi identities, let us discuss the gauge symmetries of the model. Using the results derived in the previous section for the case without scalars, one can see that the general variation of an object of the form $\Omega=\cG^{-1}\del\cG\,$, where $\cG$ has total degree zero, can be written as
\begin{equation}
\delta\Omega=\del(\cG^{-1}\delta\cG)+[\Omega,\cG^{-1}\delta\cG]\equiv\del_\Omega(\cG^{-1}\delta\cG)\;.    
\end{equation}
Given that $\cG=e^\cA e^\phi$ one has
\begin{equation}\label{g-1dg}
\begin{split}
\cG^{-1}\delta\cG &= e^{-\phi}e^{-\cA}\delta(e^\cA e^\phi)=e^{-\phi}(e^{-\cA}\delta e^\cA) e^\phi+e^{-\phi}\delta e^\phi\\
&=\cV^{-1}\Delta\cA\,\cV+\cV^{-1}\delta\cV\;,
\end{split}
\end{equation}
where the covariant variation $\Delta\cA$ was defined in \eqref{DeltaA}. Given that $\del_\Omega^2=0$ thanks to \eqref{Omegaflat}, $\Omega$ can be made gauge invariant (with respect to the $\Lambda$ gauge parameters) by choosing
\begin{equation}\label{gaugetranswithphi}
\cG^{-1}\delta_\Lambda\cG=\del_\Omega(\cV^{-1}\Lambda\cV)=\del(\cV^{-1}\Lambda\cV)+[\Omega,\cV^{-1}\Lambda\cV] \;, 
\end{equation}
where the conjugation by $\cV^{-1}$ has been chosen in view of \eqref{g-1dg}. The first term gives
\begin{equation}
\begin{split}
\del(\cV^{-1}\Lambda\cV) &= \cV^{-1}\del\Lambda\cV-[\cV^{-1}\del\cV,\cV^{-1}\Lambda\cV]\\
&=\cV^{-1}\del\Lambda\cV-[T-\Theta+P+Q,\cV^{-1}\Lambda\cV]+\cV^{-1}[\cD A_1,\Lambda]\cV\\
&=\cV^{-1}D\Lambda\cV+\cV^{-1}\cD\Lambda\cV-[T-\Theta+P+Q,\cV^{-1}\Lambda\cV]\;.
\end{split}    
\end{equation}
By summing the second term in \eqref{gaugetranswithphi} with the decomposition \eqref{Omegadecomposed} and \eqref{g-1dg} one finally obtains
\begin{equation}
\Delta_\Lambda\cA+\delta_\Lambda\cV\,\cV^{-1}=D\Lambda+\cD\Lambda+[\cF,\Lambda]\;,
\end{equation}
that decomposes as
\begin{equation}
\delta_\Lambda\cV\,\cV^{-1}=\cD\lambda_0\;,\quad    \Delta_\Lambda\cA=D\Lambda+\cD(\Lambda-\lambda_0)+[\cF,\Lambda]\;.
\end{equation}
We recall that by $D$ we denote the $A_1-$covariant derivative $D=d+[\cD A_1,\cdot]\,$, while the $Q-$covariant derivative is denoted by $D_Q=d+[Q,\cdot]\,$.
Since $\delta_\Lambda\Omega=0$ one immediately has that $P$ and $Q$ are gauge invariant under $\Lambda\,$, while for $p\geq2$ one easily finds the usual gauge transformations for the curvatures:
\begin{equation}
\begin{split}
\delta_\Lambda F_p=\delta_\Lambda(\cV\Omega_p\cV^{-1})=[\delta_\Lambda\cV\,\cV^{-1},\cV\Omega_p\cV^{-1}]=[\cD\lambda_0,F_p]=\cL_{\lambda_0}F_p  \;.  
\end{split}    
\end{equation}

Even if $\Omega$ is gauge invariant under the $\Lambda$ transformations of the tensor hierarchy, $\del_\Omega$ is an $H-$covariant derivative with respect to the local $H-$transformations of the scalar coset. A more natural split compared to \eqref{delOmegadecomposed} thus seems
\begin{equation}\label{Omegadecomposednew}
\del_\Omega=D_Q+\tilde\cF\;,\quad\tilde\cF=T+P+\cV^{-1}\cF\,\cV \;, \end{equation}
since $\tilde\cF'=h^{-1}\tilde\cF h$ under local $H-$transformations. 
The Bianchi identities in terms of $\tilde\cF$ take then the form
\begin{equation}
D_Q\tilde\cF+R(Q)+\tfrac12[\tilde\cF,\tilde\cF]=0\;.    
\end{equation}
In the $H-$invariant ``untilded'' basis one has
\begin{equation}
\cV\tilde\cF\cV^{-1}=\cF+\cV P\cV^{-1}+\Theta    \;, 
\end{equation}
and, in order to write duality relations in a homogeneous form, we define $F_1:=\cV P\cV^{-1}$ and $F_0:=\Theta\,$, and include them in the definition of $\cF$ by renaming $F_0+F_1+\cF\to\cF\,$. $\del_\Omega$ can be written in terms of the two bases as
\begin{equation}
\del_\Omega=D_Q+\tilde\cF=D_Q+\cV^{-1}\cF\cV\;.    
\end{equation}
The Bianchi identities for the new $\cF$ read
\begin{equation}\label{BianchinewF}
D\cF+\tfrac12[\cF,\cF]-[F_1,\cF]+\cV R(Q)\cV^{-1}=0  \;, 
\end{equation}
that split as $d\Theta=0$ for the one-form, and 
\begin{equation}
DF_1-\tfrac12[F_1,F_1]+\cD F_2+\cV R(Q)\cV^{-1}=0   \;, 
\end{equation}
being just the conjugation by $\cV$ of \eqref{RQ}, \eqref{PBianchi}, and the usual \eqref{Bianchicomponent} for the gauge curvatures.

The main difference between $\cF$ and $\tilde\cF$ is that $\cF$ is $H-$invariant and transforms as a tensor under the gauge transformations of the tensor hierarchy, while 
$\tilde\cF$ is gauge invariant with respect to the tensor hierarchy and a tensor under $H\,$. We will thus refer to $\cF$ and $\tilde\cF$ as the curved and flat field basis, respectively.
Having an object transforming as a tensor under all the symmetries of the model is especially important in view of writing down dynamical equations as duality relations, that will be the goal of the next section.

\section{Dynamical equations from duality relations} 

So far our construction has only provided the kinematical data of the tensor hierarchy, namely the tower of gauge covariant curvatures built from the $p-$form gauge potentials, plus the geometry of the scalar manifold. Our goal in this section is to write down duality relations between the field strengths that, by virtue of the Bianchi identities \eqref{BianchinewF}, yield as integrability conditions second order dynamical equations for the $p-$form gauge fields and the scalars.

To begin with, in order to write the Hodge star operator we introduce a spacetime metric $g_{\mu\nu}\,$, that is inert w.r.t.~the dgLa structure, and in particular has zero degree. This should 
eventually be generalized, especially in view of the fact that in exceptional field theories the metric does transform non-trivially under the generalized diffeomorphisms that in our language correspond to the transformation $\lambda_0\,$.  
Given a $p-$form $\alpha_p=\frac{1}{p!}\,\theta^{\mu_1}...\theta^{\mu_p}\,\alpha_{\mu_1...\mu_p}$ with arbitrary internal degree $d_\alpha$ and total degree $|\alpha_p|=d_\alpha-p\,$, we define the Hodge dualization in $n$ spacetime dimensions as
\begin{equation}\label{hodge}
\star\alpha_p=\tfrac{1}{p!(n-p)!}\,\theta^{\mu_1}...\theta^{\mu_{n-p}}\,\varepsilon_{\nu_1...\nu_p\mu_1...\mu_{n-p}}\,\alpha^{\nu_1...\nu_p}  \;,  
\end{equation}
where indices have been raised with the metric $g^{\mu\nu}\,$, and $\varepsilon_{\mu_1...\mu_n}=\sqrt{|g|}\,\epsilon_{\mu_1...\mu_n}$ is the covariant volume form. The dualization \eqref{hodge} can also be realized as a differential operator (in $\theta$ space) acting on the volume form $\omega:=\frac{1}{n!}\,\theta^{\mu_1}...\theta^{\mu_n}\,\varepsilon_{\mu_1...\mu_n}$ as
\begin{equation}
\star\alpha_p=\alpha_p^\dagger\,\omega\;,\quad \alpha_p^\dagger:=(-1)^{d_\alpha(n-p)}\,\tfrac{1}{p!}\,\alpha^{\mu_1...\mu_p}\frac{\del}{\del\theta^{\mu_p}}...\frac{\del}{\del\theta^{\mu_1}}   \;, 
\end{equation}
and by degree counting one can see that the Hodge star operator $\star_p$ (meaning that it acts on a $p-$form) has total degree $|\star_p|=2p-n\,$. Moreover, following from the definition \eqref{hodge} one has $\star^2=(-1)^{p(n-p)+s}\,$, where $p$ is the form degree of the object acted upon, and $s$ is zero or one for euclidean and lorentzian signatures, respectively. 
Furthermore, it is useful to introduce the covariant divergence operator
\begin{equation}
D^\dagger:=\star D\,\star  \;,  
\end{equation}
that acts on a $p-$form as 
\begin{equation}
D^\dagger\alpha_p=\tfrac{(-1)^{n(p+1)+s}}{(p-1)!}\,\theta^{\mu_1}...\theta^{\mu_{p-1}}\,D^\nu\alpha_{\nu\mu_1...\mu_{p-1}}    \;,
\end{equation}
where $D_\mu$ contains both $A_\mu$ and the Christoffel connection of $g_{\mu\nu}$.

At this point, if one tries to impose naively a duality relation of the form $F_{p+1}=\star \,F_{n-p-1}\,$, an immediate problem arises: First of all, the degrees on the two sides do not match and, second, the two curvatures take values in different spaces, namely\footnote{This is a slight abuse of notation in place of $F_{p+1}\in Z^p_{[p+1]}$ or ``taking values in $X_p$''.} $F_{p+1}\in X_p$ and $F_{n-p-1}\in X_{n-p-2}\,$, that in practice correspond to different representations of the duality group $G\,$. In order to remedy this we shall introduce more structures, that provide dynamical data not contained in the dgLa.

\subsection{Metrics, $G-$representations and dual spaces}

First of all, let us focus on the simpler case of an ungauged scalar non-linear sigma model, that is obtained from $\Omega$ as in \eqref{Omegawithphifactor} by setting $\cA=0$ and $\del=d\,$. In this case one simply has
\begin{equation}
\Omega=\cV^{-1}d\cV=P+Q    \;,
\end{equation}
with $P\in\mathfrak{p}\,$, and $Q\in\mathfrak{h}$ being the $H$ connection.
To construct an action one usually assumes the existence of an $H-$invariant bilinear form
\begin{equation}
\langle\cdot,\cdot\rangle:\mathfrak{p}\times\mathfrak{p}\rightarrow \mathbb{R}   \end{equation}
that allows to write (in lorentzian signature)
\begin{equation}
S=-\tfrac12\int d^n x\sqrt{|g|}\,\langle P_\mu,P^\mu\rangle \;.
\end{equation}
Given that $\mathfrak{p}\subset\mathfrak{g}\equiv X_0\,$, the inner product $\langle\cdot,\cdot\rangle$ can be viewed as an $H-$invariant metric on $X_0$ that can be made diagonal, \emph{i.e.} $\langle x,y\rangle=0$ if $x\in\mathfrak{h}$ and $y\in\mathfrak{p}\,$. In turn, this is equivalent to an invertible map
\begin{equation}
\Delta_0:X_0\rightarrow X_0^*  \;,  
\end{equation}
such that the action can be rewritten as
\begin{equation}
S=-\tfrac12\int d^n x\sqrt{|g|}\,(P_\mu,\Delta_0\,P^\mu)   \;, 
\end{equation}
with round brackets denoting the natural pairing between a space and its dual. At this point we shall extend this construction to the whole space by assuming the existence of symmetric invertible maps
\begin{equation}
\Delta_p: X_p\rightarrow X_p^*\;,\quad |\Delta_p|=-2p    
\end{equation}
for all $p$'s, yielding the inner products
\begin{equation}
\langle v_p,w_p\rangle:=(v_p,\Delta_pw_p)=(w_p,\Delta_p v_p)\;.
\end{equation}
One can also consider the formal sum 
\begin{equation}
\Delta:=\sum_{p}\Delta_p:X\rightarrow X^*    \;,
\end{equation}
defined by its diagonal action on a non-homogeneous element $u=\sum_pu_p\,$:
\begin{equation}
\Delta u:=\sum_p\Delta_p u_p\;,    
\end{equation}
and likewise define the pairing and inner product on the whole space by
\begin{equation}
(u,\omega):=\sum_p(u_p,\omega_p)\;,\quad u\in X\;,\; \omega\in X^*  \;,\quad\langle u,v\rangle:=(u,\Delta v) \;. 
\end{equation}

As we mentioned before, the dgLa bracket gives a natural action of the Lie algebra $\mathfrak{g}=X_0$ on all the $X_p$ spaces, making them into $\mathfrak{g}-$representations:
\begin{equation}
\rho_x\,u:=[x,u]\;,\quad x\in\mathfrak{g}\;,\;u\in X \;.   
\end{equation}
This action can be exponentiated, yielding the group representation: For $g=e^x\in G$ one has
\begin{equation}
R_g \,u:=e^{\rho_x}u=\sum_{n=0}^\infty\tfrac{1}{n!}\,[x,[x,...[x,u]]]=e^xu\,e^{-x}=g\,u\,g^{-1}    \;,
\end{equation}
and both $\rho_x$ and $R_g$ are endomorphisms on each $X_p$ separately, given that $|x|=0\,$.
Having introduced the dual complex $X^*\,$, it is possible to define the dual representation $R_g^*$ and the transpose involution by 
\begin{equation}
(R_g\,u,R^*_g\,\omega)=(u,\omega)\;,\quad (R_g\,u,\omega)=(u,R_g^{\rm T}\,\omega) \;,  \quad u\in X\;,\;\omega\in X^* 
\end{equation}
implying that $R_g^*=(R_g^{\rm T})^{-1}$. The $H-$invariance of $\Delta$ can now be stated as
\begin{equation}\label{HinvDelta}
R_h^{\rm T}\,\Delta\,R_h=\Delta\;,\quad h\in H\;,
\end{equation}
that can be derived by demanding $\langle R_hu,R_hv\rangle=\langle u,v\rangle\,$. Infinitesimally, for $\epsilon\in\mathfrak{h}\,$, this reads $\rho_\epsilon^*\Delta=\Delta\rho_\epsilon\,$.

The discussion of $\Delta$ has involved algebraic properties pertaining to the ``internal'' structure alone, being completely insensitive of any field content in spacetime. Indeed, as shown throughout the previous sections, the kinematical construction of gauge covariant curvatures for the tensor hierarchy is completely unaffected by, for instance, the spacetime dimension. In contrast, it is well known that the group $G$ and its representations carried by spacetime fields crucially depend on the spacetime dimension, in both gauged supergravity and exceptional field theory. 
This is reflected, in our more abstract setting, by the fact that the structures introduced so far are not sufficient to construct the dynamics. The extra missing ingredient, that we introduce now, is to assume the existence of an isomorphism $I=\sum_pI_p$ between $X$ and $X^*\,$, that acts non-diagonally on the complex and makes the spacetime dimension enter explicitly:
\begin{equation}
I_p:X_p^*\rightarrow{X_{n-2-p}} \;,\quad |I_p|=n-2\;.   
\end{equation}
The isomorphism $I_p$ may seem quite unnatural from our abstract point of view, but it turns out to exist in all known examples, taking different forms. For instance, in gauged supergravity and exceptional field theory (with the caveat that in the latter case the metric itself transforms non-trivially), $I_p$ is just the identity, endowed with a degree shift of $n-2$ for degree matching, since in those cases it just happens that $X_p^*\equiv X_{n-2-p}$ as vector spaces, as already discussed in \cite{Palmkvist:2011vz,Palmkvist:2013vya} . In the more subtle case of self-duality conditions (as for two-form curvatures in $n=4$), the isomorphism $I_p$ is typically provided by an extra structure, as for instance the $Sp(56)$ symplectic matrix $\Omega_{MN}$ in four dimensions. Yet another example is the gauge theory based on volume-preserving diffeomorphisms \cite{Hohm:2018git}, where the isomorphism $I_p$ is the Hodge star in the internal manifold, mapping multi-vectors  to differential forms. 
Here we would just like to stress that the existence of the maps $I_p$ gives the first constraints on the representation content of the spacetime fields.
Furthermore, the existence of such an isomorphism was suggested in \cite{Cremmer:1998px} in the context of ungauged supergravity. 

Thanks to the isomorphisms $I_p\,$, one has two different ways to map elements from $X_p$ to $X_p^*\,$, and thus two different inner products: The first map is simply given by $\Delta$
\begin{equation}
X_p\stackrel{\Delta}{\longrightarrow} X_p^*   \;,  
\end{equation}
while the second involves both $I$ and $\Delta$ in several steps, namely
\begin{equation}
X_p\stackrel{I^{-1}}{\longrightarrow}X_{n-2-p}^*\stackrel{\Delta^{-1}}{\longrightarrow}X_{n-2-p}\stackrel{I^{-1}}{\longrightarrow}X_p^* \;,   
\end{equation}
where the degrees of the different $\Delta$'s and $I$'s are determined by the space indicated. For the two maps to be identified one needs the compatibility condition $\Delta\,\propto\,(I\Delta I)^{-1}$ or, with explicit degrees,
\begin{equation}\label{IDeltacompatibility}
(I_p\Delta_p)^{-1}\,\propto\, I_{n-p-2}\Delta_{n-p-2}\;,
\end{equation}
where the proportionality, rather than equality, is due to a non-trivial sign factor. This sign will be determined below by requiring the strict equality of the maps at the level of the diagonal dgLa $Z$, that entails adding the spacetime Hodge star.
As another compatibility condition, we also require that $I$ be $G-$covariant, \emph{i.e.}
\begin{equation}\label{GcovI}
R_g\,I\omega=I\,R_g^*\omega\;,\quad \omega\in X^*\;.    
\end{equation}
In fact, $G-$covariance of $I$ is just a special case of a more general compatibility requirement that we are going to state: Given any element $u\in X_p\,$, one has a degree $p$ map $\hat u$ in the chain complex $X\,$,
\begin{equation}\label{hatoperator}
\hat u:X_k\rightarrow  X_{k+p}\;,\quad \hat u(v):=[u,v]\;.    
\end{equation}
Having introduced the dual complex $X^*\,$, the dual map $\hat u^*:X^*_{k+p}\to X^*_k$ is defined by the pairing as
\begin{equation}
\left(\hat u(v),\omega\right)=(-1)^{1+|u||v|}\left(v,\hat u^*(\omega)\right)\;,  \end{equation}
where the sign convention has been chosen as the graded generalization of the Lie algebra coadjoint action: $(\rho_x u,\omega)=-(u,\rho_x^*\omega)\,$, since for $p=0$ one has $\hat u\equiv\rho_u\,$. Thanks to the $I$ isomorphism, given $u\in X_p\,$, one can also transport the map $\hat u$ to $X^*$ by
\begin{equation}
I^{-1}\hat uI:X^*_{k+p}\rightarrow X^*_k\;.    
\end{equation}
We will demand compatibility by asking that the two maps on $X^*$ coincide:
\begin{equation}\label{Jade}
\hat u^*=I^{-1}\hat uI\quad\rightarrow\quad \left([u,v],\omega\right)=(-1)^{1+|u||v|}\left(v,I^{-1}[u,I\omega]\right)\;.    
\end{equation}

\subsection{Flat duality relations}

Given the two bases for curvatures, namely the usual $\cF$ and the ``flat'' version $\tilde\cF\,$, one can write duality relations for both, but it turns out that the flat basis makes consistency easier to prove, while the usual one gives rise to dynamical equations in the standard form. We will then discuss the duality relations in terms of $\tilde\cF$ here, and analyze them in the $\cF$ basis in the next subsection.

As we mentioned before, trying to impose relations in the naive form $\tilde F_{n-p-1}=\star\,\tilde F_{p+1}$ does not make sense both for degree counting and vector spaces not matching, meaning that one has to supplement the action of the Hodge star by an extra operator. This extra operator is precisely given by the combination $I\Delta\,$. Indeed, when acting on a curvature $\tilde F_{p+1}\in X_p$ the combination $\star I_p\Delta_p$ has total degree zero\footnote{Recall that $|I_p\Delta_p|=n-2-2p$ and $|\star_p|=2p-n$ acting on a $p-$form.}, and $\star I_p\Delta_p\,\tilde F_{p+1}$ is an $(n-p-1)-$form taking values in $X_{n-p-2}\,$, that can be equated to $F_{n-p-1}\,$. We thus impose the duality relation
\begin{equation}\label{dualityflat}
\tilde\cF=\star I\Delta\,\tilde\cF   \;, 
\end{equation}
or, in components
\begin{equation}
\tilde F_{n-p-1}=\star I_p\Delta_p\,\tilde F_{p+1}\;,\quad p=-1,0,...,n-1\;.    
\end{equation}
Consistency of the above equation requires $(\star I\Delta)^2=1$ that reads
\begin{equation}
\star_{n-p-1}I_{n-p-2}\Delta_{n-p-2}\star_{p+1}I_p\Delta_p=1
\end{equation}
with explicit degrees. By using $|I_p\Delta_p|=n-2(p+1)$ and $|\star_q|=2q-n$ one has
\begin{equation}
I_p\Delta_p\star_q=(-1)^n\star_qI_p\Delta_p\;,
\end{equation}
that, together with $\star_{n-p}\star_p=(-1)^{p(n-p)+s}$ fixes the sign in \eqref{IDeltacompatibility}:
\begin{equation}\label{IDeltacompatibilityprecise}
(I_p\Delta_p)^{-1}=(-1)^{p(n-p-2)+1+s} I_{n-p-2}\Delta_{n-p-2}\;. 
\end{equation}

Having established the consistency of $(1-\star I\Delta)$ as a projector, we now turn to study compatibility of \eqref{dualityflat} with the symmetries of the theory. Gauge symmetry under the $\Lambda$ transformations of the tensor hierarchy is trivially respected, since $\tilde \cF$ is invariant. On the other hand, $\tilde\cF$ transforms as $h^{-1}\tilde\cF\,h$ under local $H-$transformations. Covariance of the equation \eqref{dualityflat} is readily proven thanks to $H-$invariance of $\Delta$ \eqref{HinvDelta} and full $G-$covariance of $I$ \eqref{GcovI}:
\begin{equation}
\begin{split}
\star I\Delta\tilde\cF\;\rightarrow\; \star I\Delta\,h^{-1}\tilde\cF\,h&=\star I\Delta\,R_{h^{-1}}\tilde\cF\\
&=\star I\,(R_h^{\rm T})^{-1}\Delta\tilde\cF =\star I\,R_h^*\Delta\tilde\cF\\
&=\star \,R_hI\Delta\tilde\cF=h^{-1}(\star I\Delta\tilde\cF)h\;, 
\end{split}
\end{equation}
therefore proving full consistency of the proposed equation \eqref{dualityflat}. 
We will now turn to transform the consistent duality relation in the $\cF$ basis, where second order dynamical equations are recovered in a more familiar form.

\subsection{Duality relations and field equations}

Since the curvatures in the ordinary basis are given by $\cF=\cV\tilde\cF\cV^{-1}\,$, we shall perform the conjugation by $\cV$ of the duality relation \eqref{dualityflat}, yielding
\begin{equation}
\cF=\star\cV I\Delta\tilde\cF\cV^{-1}\;,    
\end{equation}
since the Hodge star commutes with the degree zero scalar $\cV\,$. We now manipulate the right hand side by using the definition of $G-$representations and $G-$covariance of the $I$ isomorphism:
\begin{equation}
\begin{split}
\cV(I\Delta\tilde\cF)\cV^{-1}&=R_\cV\,I\Delta\tilde\cF =I\,R_\cV^*\Delta\tilde\cF =I\,R_\cV^*\Delta\cV^{-1}\cF\cV =I\,R_{\cV}^*\Delta R_\cV^{-1}\cF=:I\cM\cF   \;,  
\end{split}
\end{equation}
where we have defined the scalar-dependent metric
\begin{equation}
\cM:=R_{\cV}^*\Delta R_\cV^{-1}\equiv (R_\cV^{\rm T})^{-1}\Delta R_\cV^{-1}\;.    
\end{equation}
Incidentally, in the literature it is common to call $\cV$ the matrix of the $R_{\cV}^*\rvert_{X_1}$ representation. The metric $\cM_1$ then takes the more familiar form $\cM_1=\cV\Delta_1 \cV^{\rm T}$. 
Under local $H-$transformations and global $G-$transformations $\cV(x)\to g\cV(x) h(x)\,$, so that ${R_\cV\to R_gR_\cV R_h}\,$. This allows us to determine the transformation of $\cM\,$:
\begin{equation}
\begin{split}
\cM\to R_g^*R_\cV^*R_h^*\Delta R_h^{-1}R_\cV^{-1}R_g^{-1}&= R_g^*R_\cV^*(R_h^{\rm T})^{-1}\Delta R_h^{-1}R_\cV^{-1}R_g^{-1}\\
&=R_g^*R_\cV^*\Delta R_\cV^{-1}R_g^{-1}=R_g^*\cM R_g^{-1}\;,
\end{split}    
\end{equation}
that is thus $H-$invariant  thanks to the $H-$invariance of $\Delta\,$.

Finally, the duality relations can be cast in the form
\begin{equation}\label{dualitycurved}
\cF=\star I\cM \cF \;,   
\end{equation}
or, in components,
\begin{equation}\label{dualitycurvedcomponents}
F_{n-p-1}=\star I_p\cM_p F_{p+1}\;,\quad p=-1,0,...,n-1\;.    
\end{equation}
Invariance under local $H-$transformations is now manifest, since $\cF$ is $H-$invariant. The proof of gauge covariance of the field equations $\cG:=(1-\star I\cM)\cF=0$ is easily carried out considering that $\cV^{-1}\cG\cV$ is gauge invariant:
\begin{equation}
0=\delta_\Lambda(\cV^{-1}\cG\cV)=\cV^{-1}\delta_\Lambda\cG\cV-[\cV^{-1}\delta_\Lambda\cV,\cV^{-1}\cG\cV]=\cV^{-1}\left(\delta_\Lambda\cG-[\cD\lambda_0,\cG]\right)\cV \;.   
\end{equation}

As a last remark before deriving the field equations by using the Bianchi identities, we shall show that  the $p=0$ component of \eqref{dualitycurvedcomponents} involves the current $J$ of the scalars on the r.h.s. If one considers the action for the scalars in the ungauged case, namely
\begin{equation}
S=-\tfrac12\int d^nx\sqrt{|g|}\,(P_\mu,\Delta_0 P^\mu)   \;, 
\end{equation}
it is possibe to find the conserved currents associated to the global $G-$symmetry. Under a global $G-$transformation $P$ is invariant, but if we consider a local parameter ${g(x)=1+\epsilon(x)}\,$, under which $\delta\cV=\epsilon\cV\,$, one has  $\delta P_\mu=[\cV^{-1}\del_\mu\epsilon\cV]\rvert_\mathfrak{p}\,$. By using this in the action we obtain
\begin{equation}
\begin{split}
\delta S&=-\int d^nx\sqrt{|g|}\,(\delta P_\mu,\Delta_0P^\mu)=    -\int d^nx\sqrt{|g|}\,(\cV^{-1}\del_\mu\epsilon\cV,\Delta_0P^\mu)\\
&=-\int d^nx\sqrt{|g|}\,(R_{\cV}^{-1}\del_\mu\epsilon,\Delta_0 P^\mu)=-\int d^nx\sqrt{|g|}\,(\del_\mu\epsilon,(R_\cV^{\rm T})^{-1}\Delta_0P^\mu)\\
&=-\int d^nx\sqrt{|g|}\,(\del_\mu\epsilon,R_\cV^*\Delta_0P^\mu)\;,
\end{split}
\end{equation}
where, in the first line, we used the symmetry property of $\Delta_0$ as well as the fact that, being diagonal, the projection on $\mathfrak{p}$ is redundant. 
The Noether current is thus found to be
\begin{equation}
J_\mu=R_\cV^*\Delta_0\,P_\mu  \;.  
\end{equation}
Despite appearing as an arbitrary $\mathfrak{g}^*-$valued current, $J_\mu$ clearly has only ${\rm dim}G-{\rm dim}H$ independent components, since it is given by an operator ($R_\cV^*\Delta_0$) acting on $P_\mu\,$, that takes values only along $\mathfrak{p}$.
The combination $\cM_p F_{p+1}$ at $p=0$ can thus be rewritten as
\begin{equation}
\cM_0F_1=R_\cV^*\Delta_0R_{\cV}^{-1}\cV P\cV^{-1}=R_\cV^*\Delta_0R_{\cV}^{-1}\,R_\cV P=R_\cV^*\Delta_0\,P=J    \;,
\end{equation}
proving that the duality relation \eqref{dualitycurvedcomponents} at $p=0$ is given by
\begin{equation}
F_{n-1}=\star I_0\,J    \;.
\end{equation}
We are now ready to use the first order duality equation \eqref{dualitycurved} to derive second order field equations upon acting with a covariant curl $D\,$. By using the Bianchi identity \eqref{BianchinewF} and $\star D\star=D^\dagger$ we finally obtain
\begin{equation}
D^\dagger(I_p\cM_p\,F_{p+1})+\tfrac12\sum_{k=2}^{n-p-2}\star\,[F_k,F_{n-p-k}]+(-1)^{\epsilon}\cD(I_{p-1}\cM_{p-1}\,F_p)=0\;,\quad p=1,...,n-3    \;,
\end{equation}
where $\epsilon=p(n-p)+n+s$ and we used the duality relations again in the last term. 

Since the scalar sector is mostly described using the one-form $P\,$, rather than the current $J$ or even $F_1\,$, we will derive the scalar equation in terms of $P\,$. In particular, we will show that it is variational and contains a source term due to a scalar potential.
We start from the duality relation
\begin{equation}
P=\star I_{n-2}\Delta_{n-2}\,\tilde F_{n-1}\;,    
\end{equation}
combined with the Bianchi identity for $\tilde F_{n-1}\,$, namely
\begin{equation}
D_Q\tilde F_{n-1}+\tfrac12\sum_{k=2}^{n-2}[\tilde F_k,\tilde F_{n-k}]+[P,\tilde F_{n-1}]+[T,\tilde F_n]=0  \;.  
\end{equation}
Thanks to $G-$invariance of $I$ and $H-$invariance of $\Delta$ one can commute the $H-$covariant derivative through: $D_QI\Delta=(-1)^nI\Delta D_Q\,$, yielding 
\begin{equation}
D_Q\star P=(-1)^s\,I_{n-2}\Delta_{n-2}\Big(\tfrac12\sum_{k=2}^{n-2}[\tilde F_k,\tilde F_{n-k}]+[P,\tilde F_{n-1}]+[T,\tilde F_n]\Big) \;.   
\end{equation}
By further using the duality relation in the last two terms and \eqref{IDeltacompatibilityprecise} one finally obtains
\begin{equation}\label{scalarequation}
D_Q\star P+(I_0\Delta_0)^{-1}\Big([P,\star I_0\Delta_0P]+[T,\star I_{-1}\Delta_{-1}T]+\tfrac12[\tilde F,\tilde F]\Big)    =0\;,
\end{equation}
where we schematically denoted the sum over terms containing gauge fields. In odd dimensions $n=2m+1$ one can use the duality relations to rewrite
\begin{equation}
\tfrac12[\tilde F,\tilde F]=\sum_{k=2}^m[\tilde F_k,\star I_{k-1}\Delta_{k-1}\tilde F_k]   \;, 
\end{equation}
while in even dimensions ${n=2m+2}$ one has to separate the contribution from the self-dual curvature $\tilde F_{m+1}\,$:
\begin{equation}
\tfrac12[\tilde F,\tilde F]=\sum_{k=2}^{m}[\tilde F_k,\star I_{k-1}\Delta_{k-1}\tilde F_k]+\tfrac12[\tilde F_{m+1},\star I_{m}\Delta_{m}\tilde F_{m+1}]   \;.
\end{equation}
We will now show that the above contributions can be obtained by varying the generalized Yang-Mills actions
\begin{equation}\label{YMs}
S_p=\tfrac{(-1)^{np+s}}{2}\int( F_{p+1},\star\cM_p F_{p+1})=\tfrac{(-1)^{np+s}}{2}\int( \tilde F_{p+1},\star\Delta_p \tilde F_{p+1})\;,
\end{equation}
except for the self-dual case in even dimensions, where it is obtained from the pseudo-action
\begin{equation}
S_m=\tfrac{(-1)^{s}}{4}\int( F_{m+1},\star\cM_m F_{m+1})\;,\quad (1-\star I_m\cM_m)F_{m+1}=0\;,   
\end{equation}
and the unusual phase factors in the definition \eqref{YMs} arise from commuting the $\theta$ oscillators in the inner product. The actions \eqref{YMs} contain, besides the proper Yang-Mills terms, the scalar kinetic term for $p=0\,$:
\begin{equation}\label{S0}
S_0=\tfrac{(-1)^{s}}{2}\int(P,\star\Delta_0 P)   \;, 
\end{equation}
and the scalar potential contribution for $p=-1$
\begin{equation}
S_{-1}=\tfrac{(-1)^{n+s}}{2}\int(T,\star\Delta_{-1} T)=\tfrac{(-1)^{s}}{2}\int d^nx\sqrt{|g|}\,(T,\Delta_{-1} T) \;.   
\end{equation}
Let us recall that the information encoded in the scalar potential, defined by
\begin{equation}
V=\tfrac12\,(T,\Delta_{-1} T)\;,    
\end{equation}
is equivalent to that in $\Delta_{-1}$. At this stage $\Delta_{-1}$ is only constrained by invertibility, symmetry and $H-$invariance, which leaves room for several possibilities, among which there is the one fixed by the particular gauged supergravity model.

In order to vary the above actions with respect to the scalar fields, one takes the Lie algebra valued variation $\cV^{-1}\delta\cV$ and projects it on $\mathfrak{p}\,$:
\begin{equation}
\cV^{-1}\delta\cV=\Delta\phi+\tau\;,\quad \Delta\phi\in\mathfrak{p}\;,\,\tau\in\mathfrak{h}\;.
\end{equation}
The variation of $P$ then reads
\begin{equation}
\delta P=D_Q\Delta\phi+[P,\tau]+[P,\Delta\phi]\rvert_\mathfrak{p}\;.    
\end{equation}
In varying the kinetic term \eqref{S0} the $[P,\tau]$ part drops out thanks to $H-$invariance of $\Delta_0\,$, and the explicit projection on $\mathfrak{p}$ is ensured by diagonality, yielding
\begin{equation}
\begin{split}
\delta_\phi S_0&=(-1)^s\int(D_Q\Delta\phi+[P,\Delta\phi],\star\Delta_0P)= (-1)^{s+1}\int\Big\{(\Delta\phi,\Delta_0D_Q\star P)-(\rho_P\Delta\phi,\star\Delta_0P)\Big\} \\
&=(-1)^{s+1}\int\big(\Delta\phi,\Delta_0\{D_Q\star P+\Delta_0^{-1}\rho_P^*\Delta_0\star P\}\big)\;.
\end{split}
\end{equation}
The contribution from the scalar kinetic term can be further manipulated by using
\eqref{Jade}, yielding
\begin{equation}
\begin{split}\label{EulerLagrange1}
D_Q\star P+\Delta_0^{-1}\rho_P^*\Delta_0\star P &= D_Q\star P+(-1)^n\Delta_0^{-1}I_0^{-1}\rho_PI_0\Delta_0\star P\\
&=D_Q\star P+(I_0\Delta_0)^{-1}[P,\star I_0\Delta_0P]\;,
\end{split}
\end{equation}
where we used $\star\Delta=\Delta\star\,$, as well as $\star I=(-1)^n I\star$ and
\begin{equation}\label{movingtheta}
    \rho_P^*=\theta^\mu\rho_{P_\mu}^*=\theta^\mu I_0^{-1}\rho_{P_\mu}I_0=(-1)^nI_0^{-1}\rho_{P}I_0\;.
\end{equation}
The contribution \eqref{EulerLagrange1} coincides with the corresponding one in \eqref{scalarequation}. The variation of the Yang-Mills terms \eqref{YMs} (including $S_{-1}$) is most easily determined in the flat basis $\tilde F_{p+1}\,$. First of all one has, for $p\neq0\,$, 
\begin{equation}
\delta_\phi\tilde F_{p+1}=\delta_\phi(\cV^{-1}F_{p+1}\cV)=[\tilde F_{p+1},\Delta\phi]+[\tilde F_{p+1},\tau]\;.    
\end{equation}
As in the previous case, the $\tau-$dependent part of the variation drops out thanks to $H-$invariance, and one obtains
\begin{equation}
\begin{split}
\delta_\phi S_p=(-1)^{np+s}\int([\tilde F_{p+1},\Delta\phi],\star\Delta_p\tilde F_{p+1})=(-1)^{np+s+1}\int(\Delta\phi,\widehat{\tilde{F}}_{p+1}^*\,\star \Delta_p\tilde F_{p+1})    \;, 
\end{split}
\end{equation}
with the hat referring to the notation introduced in (\ref{hatoperator}). 
The corresponding contribution to the scalar equation can thus be written as
\begin{equation}
\begin{split}
(-1)^{np}\Delta_0^{-1}\widehat{\tilde{F}}_{p+1}^*\,\star \Delta_p\tilde F_{p+1}&=(-1)^n\Delta_0^{-1}I_0^{-1}[\tilde F_{p+1},I_p\star\Delta_p\tilde F_{p+1}]    \\
&=(I_0\Delta_0)^{-1}[\tilde F_{p+1},\star I_p\Delta_p\tilde F_{p+1}]\;,
\end{split}    
\end{equation}
where we used again \eqref{Jade} and the manipulation analogous to \eqref{movingtheta}, thus confirming that the scalar equation \eqref{scalarequation} coincides, including the contribution from the scalar potential, with the variational one obtained from $\sum_p S_p\,$.

In fact, a similar computation shows that if one introduces a Yang-Mills action for all $p\,$, both the scalar and gauge field equations are compatible with the single pseudo action
\begin{equation}\label{democraticaction}
S=\tfrac14\,\int(\cF,\star\cM\cF)\;,\quad (1-\star I\cM)\cF=0\;.    
\end{equation}
The field equations are already implied by the duality relations, but the pseudo action can still be useful in the context of supergravity, since by adding the suitable Einstein-Hilbert term it provides the gauge field contributions to the stress energy tensor.

\section{Outlook}

In this paper we have discussed  differential graded Lie algebras as a universal algebraic structure allowing for the construction  
of tensor hierarchies and the formulation of duality relations. The duality relations encode dynamics in that the second-order field equations 
follow as integrability conditions from the first-order duality relations together with the Bianchi identities of the tensor hierarchy. 
It would be interesting  to extend this research in the following directions: 
\begin{itemize}
\item We have treated the (external) spacetime metric $g_{\mu\nu}$ as fixed, and so it would be important 
to include a dynamical spacetime metric. In the context of gauged supergravity this is straightforward as the diffeomorphisms 
do not mix with the gauge symmetries of the tensor hierarchy. Accordingly, the complete bosonic dynamics can be encoded by  
a pseudo-action that adds the Einstein-Hilbert term to (\ref{democraticaction}), while the duality relations are unchanged. 
In the more general exceptional field theories the situation is more subtle (and more intriguing) for here the external metric 
transforms non-trivially under internal generalized diffeomorphisms as  part of the tensor hierarchy. 
One may then ask whether, in the present scheme, there is a natural place for a dynamical metric and, 
in particular, whether its dynamics can be encoded in first-order duality relations, see \cite{Hohm:2018qhd,Boulanger:2008nd}.

\item Another natural question is whether it is possible to include  mixed Young tableaux fields via exotic dualities as in the recent paper \cite{Chatzistavrakidis:2019len}
whose formalism shares some key features with our approach here.

\item Arguably one of the most important open problems in exceptional field theory is to find a universal formulation `without split', i.e., one for which 
there is no a priori split into `external' and `internal' (generalized) spacetimes. The results presented here suggest a natural strategy: finding a 
dgLa as in (\ref{introComplex}) whose brackets and differential $\partial$ are defined in some intrinsic fashion (perhaps as in \cite{Cederwall:2019qnw,Cederwall:2019bai}), 
rather than being derived by tensoring a given smaller dgLa with forms of a fixed spacetime manifold. Such an algebra should then give back 
the present formulation upon suitable  `level-decompositions' with respect to which the differential decomposes as $\partial=d+\cD$, with the 
spacetime de Rham differential $d$ and the internal differential $\cD$ that is covariant under the U-duality group in the dimension considered. 
(See \cite{Bossard:2019ksx} for recent related results.)

\end{itemize}

\subsection*{Acknowledgements}
We would like to thank Jakob Palmkvist for correspondence and useful explanations of \cite{Greitz:2013pua}. 

This work is supported by the ERC Consolidator Grant ``Symmetries \& Cosmology".

\appendix

\section{Field redefinition of gauge $p-$forms}\label{BCH section}

In this appendix we will provide the field redefinition between the field basis used in the present paper and the one employed in \cite{Bonezzi:2019ygf}, that singles out the vector $A_1$ and makes direct contact with usual expressions in the literature.
In this section we will use both pictures of the enhanced Leibniz algebra and dgLa. Since the two can be distinguished by the presence of either the bullet product $\bullet$ or the bracket $[\;,\,]\,$, we will not use different names, or tildes, for fields in order not to clutter the expressions.

In \cite{Bonezzi:2019ygf} it was proven that gauge covariant curvatures, grouped in the formal sum ${\cF:=\sum_{p=2}^\infty F_p}\,$, can be defined as
\begin{equation}\label{oldFs}
\cF=\sum_{N=0}^\infty\frac{(-\iota_{\mathbf{A}})^N}{N!}\Big[\tfrac{1}{N+1}\,(D+\cD)\mathbf{A}+{\omega}\Big] \;,  \quad \iota_\mathbf{A}:=\mathbf{A}\bullet\;, 
\end{equation}
where $\mathbf{A}:=\sum_{p=2}^\infty A_p\,$ is the formal sum of higher form gauge fields, while ${\omega}:=\sum_{p=2}^\infty\omega_p\,$ contains pseudo Chern-Simons forms built out of $A_1$ as 
\begin{equation}
\omega_p=\tfrac{(-1)^p}{(p-1)!}\,\iota_{A_1}^{p-2}\Big(dA_1-\tfrac1p\,A_1\circ A_1\Big) \;,
\end{equation}
and the covariant derivative is given by $D=d-\cL_{A_1}\,$.
The curvatures \eqref{oldFs} obey the Bianchi identities
\begin{equation}\label{oldBianchi}
D\cF+\tfrac12\,\cF\bullet\cF=\cD\cF\;,    
\end{equation}
that were used to prove gauge covariance $\delta_\lambda\cF=\cL_{\lambda_0}\cF$ recursively.
Thanks to the new definitions of the Lie derivative \eqref{LieLeibnew} and \eqref{LiedgLa}, the covariant derivative takes the form 
\begin{equation}\label{newD}
D_\mu a:=\del_\mu a-\cL_{A_\mu}a=\del_\mu a+\cD A_\mu\bullet a \;, \end{equation}
making it possible to rewrite the curvatures $F_p$ in a way that does not single out $A_1$ as a special field. We start by writing down the first curvatures from \eqref{oldFs} with the $\omega_p$ forms explicit:
\begin{equation}\label{firstFs}
\begin{split}
F_2&= dA_1-\tfrac12\,A_1\circ A_1+\cD A_2\;,\\
F_3&= DA_2+\Omega_3+\cD A_3=DA_2-\tfrac12\,A_1\bullet dA_1+\tfrac16\,A_1\bullet(A_1\circ A_1)+\cD A_3\;,\\
F_4&=DA_3+\Omega_4-A_2\bullet\Omega_2-\tfrac12\,A_2\bullet\cD A_2+\cD A_4\\
&=DA_3+\tfrac16\,A_1\bullet(A_1\bullet dA_1)-\tfrac{1}{24}\,A_1\bullet(A_1\bullet(A_1\circ A_1))-A_2\bullet dA_1+\tfrac12\,A_2\bullet(A_1\circ A_1)\\
&-\tfrac12\,A_2\bullet\cD A_2+\cD A_4\;.
\end{split}    
\end{equation}
Thanks to \eqref{Leibniz defined} and \eqref{newD}, these can be rewritten in the more symmetric form
\begin{equation}\label{firstFsnew}
\begin{split}
F_2&=dA_1+\cD A_2-\tfrac12\,A_1\bullet\cD A_1\;,\\ 
F_3&=dA_2+\cD A'_3-\tfrac12\,A_1\bullet dA_1-\tfrac12\,[A_2\bullet\cD A_1+A_1\bullet\cD A_2]+\tfrac16\,A_1\bullet(A_1\bullet \cD A_1)\;,\\
F_4&=dA'_3+\cD A'_4-\tfrac12\,[A_2\bullet dA_1+A_1\bullet dA_2]-\tfrac12\,[A'_3\bullet\cD A_1+A_2\bullet\cD A_2+A_1\bullet\cD A'_3]\\
&+\tfrac16\,A_1\bullet(A_1\bullet dA_1)+\tfrac16\,[A_2\bullet(A_1\bullet\cD A_1)+A_1\bullet(A_1\bullet\cD A_2)+A_1\bullet(A_2\bullet\cD A_1)]\\
&-\tfrac{1}{24}\,A_1\bullet(A_1\bullet(A_1\bullet\cD A_1))
\end{split}    
\end{equation}
upon performing the field redefinitions
\begin{equation}
A_3':=A_3-\tfrac12\,A_1\bullet A_2\;,\quad A_4':=A_4-\tfrac12\,A_1\bullet A_3'-\tfrac16\,A_1\bullet(A_1\bullet A_2)  \;.  
\end{equation}
This suggests that, upon defining 
\begin{equation}
\cA':=\sum_{p=1}^\infty A_p'   
\end{equation}
with $A_1'\equiv A_1$ and $A_2'\equiv A_2\,$,
it is possible to recast the entire set of curvatures in the form
\begin{equation}\label{newF'}
\Omega:=\cF+\cD A_1=\sum_{N=0}^\infty\frac{(-\iota_{\cA'})^N}{(N+1)!}(d+\cD)\cA' \;.   
\end{equation}
Upon degree shifting of both gauge fields and curvatures, the field strengths in \eqref{firstFsnew} become
\begin{equation}
\begin{split}
 F_2&=d A_1+\cD A_2+\tfrac12\,[\cD A_1, A_1]\;,\\
F_3&=d A_2+\cD A'_3+\tfrac12\,[d A_1, A_1]+\tfrac12\,\big\{[\cD A_1, A_2]+[\cD A_2, A_1]\big\}+\tfrac16\,[[\cD A_1, A_1], A_1]\;,\\
 F_4&=d A'_3+\cD A'_4+\tfrac12\,\big\{[d A_1, A_2]+[d A_2, A_1]\big\}+\tfrac12\,\big\{[\cD A_1, A'_3]+[\cD A_2, A_2]+[\cD A'_3, A_1]\big\}\\
&+\tfrac16\,[[d A_1, A_1], A_1]+\tfrac16\,\big\{[[\cD A_2, A_1], A_1]+[[\cD A_1, A_2], A_1]+[[\cD A_1, A_1], A_2]\big\}\\
&+\tfrac{1}{24}\,[[[\cD A_1, A_1], A_1], A_1]\;,
\end{split}    
\end{equation}
that indeed coincide with the expansion of \eqref{OmegaMCdefined}
\begin{equation}
\Omega=\cF+\cD A_1=\del\cA'+\tfrac12\,[\del\cA',\cA']+\tfrac16\,[[\del\cA',\cA'],\cA']+\tfrac{1}{24}\,[[[\del\cA',\cA'],\cA'],\cA']+...    
\end{equation}
In the dgLa picture it is possible to determine the field redefinition $\mathbf{A}'(\mathbf{A},A_1)$ to all orders. Given
$\cA'=A_1+\mathbf{A}'$ one uses the Baker-Campbell-Hausdorff field redefinition
\begin{equation}
\mathbf{A}'=\ln\left(e^{A_1}e^{\mathbf{A}}\right) -A_1   
\end{equation}
and obtain
\begin{equation}
\begin{split}
\Omega&=e^{-\cA'}\del\,e^{\cA'}=e^{-A_1-\mathbf{A}'}\del\,e^{A_1+\mathbf{A}'}=e^{-\mathbf{A}}e^{-A_1}\del\left(e^{A_1}e^{\mathbf{A}}\right)  \\
&=e^{-\mathbf{A}}\del\,e^{\mathbf{A}}+e^{-\mathbf{A}}\left(e^{-A_1}\del\,e^{A_1}\right)e^{\mathbf{A}}\\
&=e^{-\mathbf{A}}\del\,e^{\mathbf{A}}+e^{\iota_{\mathbf{A}}}(\omega+\cD A_1)\\
&=\cD A_1+\sum_{N=1}^\infty\frac{(\iota_{\mathbf{A}})^{N-1}}{N!}(D+\cD)\mathbf{A}+e^{\iota_{\mathbf{A}}}\omega\;,
\end{split}    
\end{equation}
that, upon suspension, coincides with \eqref{oldFs}.


\begin{thebibliography}{99}

%\cite{deWit:2002vt,deWit:2004nw,deWit:2005hv}


\bibitem{deWit:2002vt}
B.~de~Wit, H.~Samtleben, and M.~Trigiante, {\it On {L}agrangians and gaugings
  of maximal supergravities},  {\em Nucl. Phys.} {\bf B655} (2003) 93--126,
  [hep-th/0212239].
  
%\cite{deWit:2004nw,deWit:2005hv}
\bibitem{deWit:2004nw} 
  B.~de Wit, H.~Samtleben and M.~Trigiante,
  ``The Maximal D=5 supergravities,''
  Nucl.\ Phys.\ B {\bf 716}, 215 (2005)
  doi:10.1016/j.nuclphysb.2005.03.032
  [hep-th/0412173].  
  
%\cite{deWit:2005hv}
\bibitem{deWit:2005hv} 
  B.~de Wit and H.~Samtleben,
  ``Gauged maximal supergravities and hierarchies of nonAbelian vector-tensor systems,''
  Fortsch.\ Phys.\  {\bf 53}, 442 (2005)
  doi:10.1002/prop.200510202
  [hep-th/0501243].  
  
%\cite{deWit:2008ta}
\bibitem{deWit:2008ta} 
  B.~de Wit, H.~Nicolai and H.~Samtleben,
  ``Gauged Supergravities, Tensor Hierarchies, and M-Theory,''
  JHEP {\bf 0802}, 044 (2008)
  doi:10.1088/1126-6708/2008/02/044
  [arXiv:0801.1294 [hep-th]].  

%\cite{LODAY,Strobl1,Hohm:2018ybo}

\bibitem{LODAY}
J.-L. Loday, {\em Cyclic homology}, vol.~301 of {\em Grundlehren der
  Mathematischen Wissenschaften [Fundamental Principles of Mathematical
  Sciences]}.
\newblock Springer-Verlag, Berlin, 1992.



\bibitem{Strobl1}  
Thomas Strobl, ``Mathematics around Lie 2-algebroids and the tensor hierarchy in gauged supergravity," 
talk at ``Higher Lie theory", University of Luxembourg, 2013.


%\cite{Hohm:2018ybo}
\bibitem{Hohm:2018ybo} 
  O.~Hohm and H.~Samtleben,
  ``Leibniz-Chern-Simons Theory and Phases of Exceptional Field Theory,''
  Commun.~Math.~Phys.~(2019), 
  doi:10.1007/s00220-019-03347-1
  arXiv:1805.03220 [hep-th].


%\cite{Kotov:2018vcz}
\bibitem{Kotov:2018vcz} 
  A.~Kotov and T.~Strobl,
  ``The Embedding Tensor, Leibniz-Loday Algebras, and Their Higher Gauge Theories,''
  arXiv:1812.08611 [hep-th].  


%\cite{Lavau:2017tvi}
\bibitem{Lavau:2017tvi} 
  S.~Lavau,
  ``Tensor hierarchies and Leibniz algebras,''
  J.\ Geom.\ Phys.\  {\bf 144}, 147 (2019)
  doi:10.1016/j.geomphys.2019.05.014
  [arXiv:1708.07068 [hep-th]].

%\cite{Bergshoeff:2009ph}
\bibitem{Bergshoeff:2009ph} 
  E.~A.~Bergshoeff, J.~Hartong, O.~Hohm, M.~Huebscher and T.~Ortin,
  ``Gauge Theories, Duality Relations and the Tensor Hierarchy,''
  JHEP {\bf 0904}, 123 (2009)
  doi:10.1088/1126-6708/2009/04/123
  [arXiv:0901.2054 [hep-th]].



\bibitem{Hohm:2013pua} 
  O.~Hohm and H.~Samtleben,
  ``Exceptional Form of D=11 Supergravity,''
  Phys.\ Rev.\ Lett.\  {\bf 111}, 231601 (2013)
  [arXiv:1308.1673 [hep-th]].  
  
%\cite{Hohm:2013vpa}
\bibitem{Hohm:2013vpa} 
  O.~Hohm and H.~Samtleben,
  ``Exceptional Field Theory I: $E_{6(6)}$ covariant Form of M-Theory and Type IIB,''
  Phys.\ Rev.\ D {\bf 89}, no. 6, 066016 (2014)
  [arXiv:1312.0614 [hep-th]].  

%\cite{Hohm:2013uia}
\bibitem{Hohm:2013uia} 
  O.~Hohm and H.~Samtleben,
  ``Exceptional field theory. II. E$_{7(7)}$,''
  Phys.\ Rev.\ D {\bf 89}, 066017 (2014)
  [arXiv:1312.4542 [hep-th]].
  
%\cite{Hohm:2014fxa}
\bibitem{Hohm:2014fxa} 
  O.~Hohm and H.~Samtleben,
  ``Exceptional field theory. III. E$_{8(8)}$,''
  Phys.\ Rev.\ D {\bf 90}, 066002 (2014)
  [arXiv:1406.3348 [hep-th]].  
  
%\cite{Abzalov:2015ega}
\bibitem{Abzalov:2015ega} 
  A.~Abzalov, I.~Bakhmatov and E.~T.~Musaev,
  ``Exceptional field theory: $SO(5,5)$,''
  JHEP {\bf 1506}, 088 (2015)
  [arXiv:1504.01523 [hep-th]].  
  
%\cite{Musaev:2015ces}
\bibitem{Musaev:2015ces} 
  E.~T.~Musaev,
  ``Exceptional field theory: $SL(5)$,''
  JHEP {\bf 1602}, 012 (2016)
  [arXiv:1512.02163 [hep-th]].  
  
%\cite{Hohm:2015xna}
\bibitem{Hohm:2015xna} 
  O.~Hohm and Y.~N.~Wang,
  ``Tensor hierarchy and generalized Cartan calculus in SL(3) × SL(2) exceptional field theory,''
  JHEP {\bf 1504}, 050 (2015)
  [arXiv:1501.01600 [hep-th]].  
  
 %\cite{Berman:2015rcc}
\bibitem{Berman:2015rcc} 
  D.~S.~Berman, C.~D.~A.~Blair, E.~Malek and F.~J.~Rudolph,
  ``An action for F-theory: $\mathrm{SL}(2)\times {{\mathbb{R}}}^{+}$ exceptional field theory,''
  Class.\ Quant.\ Grav.\  {\bf 33}, no. 19, 195009 (2016)
  [arXiv:1512.06115 [hep-th]]. 
  
%\cite{Bonezzi:2019ygf}
\bibitem{Bonezzi:2019ygf}
  R.~Bonezzi and O.~Hohm,
  ``Leibniz Gauge Theories and Infinity Structures,''
  arXiv:1904.11036 [hep-th].

%\cite{Lavau:2019oja}
\bibitem{Lavau:2019oja} 
  S.~Lavau and J.~Palmkvist,
  ``Infinity-enhancing of Leibniz algebras,''
  arXiv:1907.05752 [hep-th].

%\cite{Cremmer:1997ct,Cremmer:1998px}
\bibitem{Cremmer:1997ct} 
  E.~Cremmer, B.~Julia, H.~Lu and C.~N.~Pope,
  ``Dualization of dualities. 1.,''
  Nucl.\ Phys.\ B {\bf 523}, 73 (1998)
  doi:10.1016/S0550-3213(98)00136-9
  [hep-th/9710119].
  
%\cite{Cremmer:1998px}
\bibitem{Cremmer:1998px} 
  E.~Cremmer, B.~Julia, H.~Lu and C.~N.~Pope,
  ``Dualization of dualities. 2. Twisted self-duality of doubled fields, and superdualities,''
  Nucl.\ Phys.\ B {\bf 535}, 242 (1998)
  doi:10.1016/S0550-3213(98)00552-5
  [hep-th/9806106].
  
%\cite{Greitz:2013pua}
\bibitem{Greitz:2013pua} 
  J.~Greitz, P.~Howe and J.~Palmkvist,
  ``The tensor hierarchy simplified,''
  Class.\ Quant.\ Grav.\  {\bf 31}, 087001 (2014)
  doi:10.1088/0264-9381/31/8/087001
  [arXiv:1308.4972 [hep-th]].
  
%\cite{Zwiebach:1992ie,Lada,Hohm:2017pnh}
\bibitem{Zwiebach:1992ie} 
  B.~Zwiebach,
  ``Closed string field theory: Quantum action and the B-V master equation,''
  Nucl.\ Phys.\ B {\bf 390}, 33 (1993)
  doi:10.1016/0550-3213(93)90388-6
  [hep-th/9206084].

%\cite{Lada:1992wc}
\bibitem{Lada:1992wc} 
  T.~Lada and J.~Stasheff,
  ``Introduction to SH Lie algebras for physicists,''
  Int.\ J.\ Theor.\ Phys.\  {\bf 32}, 1087 (1993)
  doi:10.1007/BF00671791
  [hep-th/9209099].

 \bibitem{Lada}
T.~Lada and M.~Markl,
``Strongly homotopy Lie algebras,"
Communications in Algebra (1994) 23
[arXiv:9406095]



%\cite{Hohm:2017pnh}
\bibitem{Hohm:2017pnh} 
  O.~Hohm and B.~Zwiebach,
  ``$L_{\infty}$ Algebras and Field Theory,''
  Fortsch.\ Phys.\  {\bf 65}, no. 3-4, 1700014 (2017)
  doi:10.1002/prop.201700014
  [arXiv:1701.08824 [hep-th]].
  
%\cite{Voronov,Getzler}  
\bibitem{Voronov}
T.~Voronov, 
``Higher derived brackets and homotopy algebras,"
[math/0304038]. 


\bibitem{Getzler}
E.~Getzler, 
``Higher derived brackets,"
[arXiv:1010.5859 [math-ph]]. 


 %\cite{Hohm:2019wql}
\bibitem{Hohm:2019wql} 
  O.~Hohm and H.~Samtleben,
  ``Higher Gauge Structures in Double and Exceptional Field Theory,''
  Fortsch.\ Phys.\  {\bf 67}, no. 8-9, 1910008 (2019)
  doi:10.1002/prop.201910008
  [arXiv:1903.02821 [hep-th]]. 

%\cite{Palmkvist:2011vz}
\bibitem{Palmkvist:2011vz}
  J.~Palmkvist,
  ``Tensor hierarchies, Borcherds algebras and E11,''
  JHEP {\bf 1202} (2012) 066
  doi:10.1007/JHEP02(2012)066
  [arXiv:1110.4892 [hep-th]].  


%\cite{Palmkvist:2013vya}
\bibitem{Palmkvist:2013vya} 
  J.~Palmkvist,
  ``The tensor hierarchy algebra,''
  J.\ Math.\ Phys.\  {\bf 55}, 011701 (2014)
  doi:10.1063/1.4858335
  [arXiv:1305.0018 [hep-th]].  
  
%\cite{Hohm:2018git}
\bibitem{Hohm:2018git} 
  O.~Hohm and H.~Samtleben,
  ``Reviving 3D ${\cal N}=8$ superconformal field theories,''
  JHEP {\bf 1904}, 047 (2019)
  doi:10.1007/JHEP04(2019)047
  [arXiv:1810.12311 [hep-th]].


%\cite{Hohm:2018qhd}
\bibitem{Hohm:2018qhd} 
  O.~Hohm and H.~Samtleben,
  ``The dual graviton in duality covariant theories,''
  Fortsch.\ Phys.\  {\bf 67}, no. 5, 1900021 (2019)
  doi:10.1002/prop.201900021
  [arXiv:1807.07150 [hep-th]].  
  
%\cite{Boulanger:2008nd}
\bibitem{Boulanger:2008nd} 
  N.~Boulanger and O.~Hohm,
  ``Non-linear parent action and dual gravity,''
  Phys.\ Rev.\ D {\bf 78}, 064027 (2008)
  doi:10.1103/PhysRevD.78.064027
  [arXiv:0806.2775 [hep-th]].  
  
%\cite{Chatzistavrakidis:2019len}
\bibitem{Chatzistavrakidis:2019len} 
  A.~Chatzistavrakidis, G.~Karagiannis and P.~Schupp,
  ``A unified approach to standard and exotic dualizations through graded geometry,''
  arXiv:1908.11663 [hep-th].  

%\cite{Cederwall:2019qnw}
\bibitem{Cederwall:2019qnw} 
  M.~Cederwall and J.~Palmkvist,
  ``Tensor hierarchy algebras and extended geometry I: Construction of the algebra,''
  arXiv:1908.08695 [hep-th].  
  
%\cite{Cederwall:2019bai}
\bibitem{Cederwall:2019bai} 
  M.~Cederwall and J.~Palmkvist,
  ``Tensor hierarchy algebras and extended geometry II: Gauge structure and dynamics,''
  arXiv:1908.08696 [hep-th].  
  
  
%\cite{Bossard:2019ksx}
\bibitem{Bossard:2019ksx} 
  G.~Bossard, A.~Kleinschmidt and E.~Sezgin,
  ``On supersymmetric E$_{11}$ exceptional field theory,''
  arXiv:1907.02080 [hep-th].  
  
\end{thebibliography}
\end{document}